\pdfoutput=1

\PassOptionsToPackage{table,xcdraw}{xcolor}
\documentclass[letterpaper,twocolumn,10pt]{article}
\usepackage{usenix2019,epsfig,endnotes}
\usepackage[table,xcdraw]{xcolor}
\usepackage{multirow}
\usepackage{graphicx}
\usepackage{subcaption}
\usepackage{pifont}
\usepackage{upgreek}
\usepackage{url}
\usepackage{amsmath}
\usepackage{algorithm2e}
\usepackage{algpseudocode}

\begin{document}

\renewcommand{\paragraph}[1]{{\bf #1}}

\title{\Large \bf Expanding across time to deliver bandwidth efficiency and low latency}

\author{
{\rm William M. Mellette, Rajdeep Das, Yibo Guo, Rob McGuinness, Alex C. Snoeren, and George Porter}\\
University of California San Diego   {\rm \lbrack Technical Report - March 28, 2019\rbrack} \\
https://circuit-switching.sysnet.ucsd.edu/
} 
\maketitle

\subsection*{Abstract}

Datacenters need networks that support both low-latency and
high-bandwidth packet delivery to meet the stringent requirements of
modern applications.
We present Opera, a dynamic network that delivers latency-sensitive
traffic quickly by relying on multi-hop forwarding in the same way as
expander-graph-based approaches, but provides near-optimal bandwidth
for bulk flows through direct forwarding over time-varying
source-to-destination circuits.  The key to Opera's design is the
rapid and deterministic reconfiguration of the network,
piece-by-piece, such that at any moment in time the network implements
an expander graph, yet, integrated across time, the network provides
bandwidth-efficient single-hop paths between all racks.  We show that
Opera supports low-latency traffic with
flow completion times comparable to cost-equivalent static topologies,
while delivering up to 4$\times$ the bandwidth for all-to-all traffic
and supporting 60\% higher load for published datacenter workloads.

\section{Introduction}

Datacenter networks are tasked with providing connectivity between an
ever-increasing number of end hosts whose link rates improve by
orders of magnitude every few years.  Preserving the ``big-switch''
illusion of full bisection bandwidth~\cite{dcswitch:sigcomm08,vl2} by
augmenting the internal switching capacity of the network accordingly
is increasingly cost prohibitive and likely soon
infeasible~\cite{ptree:hotnets16}.
Practitioners have long favored
over-subscribed networks that provide all-to-all connectivity, but
at only a fraction of host-link speeds~\cite{vl2,jupiter}.  Such
networks realize cost savings by dramatically reducing the amount of
in-network capacity (in terms of both the number and rate of links and
switches internal to the network fabric), providing
full-speed connectivity between only a subset of hosts, and
more limited capacity between others.

The catch, of course, is that any under-provisioned topology
inherently biases the network toward certain workloads.  Traditional
over-subscribed Clos topologies only support rack-local traffic at
full line rate; researchers have proposed alternate ways of deploying
a limited amount of switching capacity---either through disparate link
and switch technologies~\cite{reactor:nsdi14,mordia:sigcomm13,shoal,cthrough},
non-hierarchical
topologies~\cite{expander1,expander2,jellyfish,xpander}, or
both~\cite{projector,rotornet:sigcomm17}---that can deliver higher
performance for published workloads~\cite{dctcp,facebook:sigcomm15} at
similar costs.
Because workloads can be dynamic, many of these proposals implement
reconfigurable networks that allocate link capacity in a time-varying
fashion, either on a fixed schedule~\cite{rotornet:sigcomm17,shoal} or in
response to recent demand~\cite{projector,reactor:nsdi14,cthrough}.
Unfortunately, practical reconfigurable technologies require
non-trivial delay to retarget capacity, limiting their utility for
workloads with stringent latency requirements.

Under-provisioned networks often incorporate some flavor of indirect
traffic routing to address inopportune traffic demands; because
application workloads do not always align well with the structure of the
network, some traffic may transit longer, less-efficient paths.
The benefits of indirection come at significant cost, however:
traversing more than a single hop through the network imposes a
``bandwidth tax.''  Said another way, $x$ bytes sent over a direct
link between two end points consume only $x$ bytes of network
capacity.  If that same traffic is instead sent over $k$ links,
perhaps indirecting through multiple switches, it consumes $(k\cdot
x)$ bytes of network capacity, where $(k-1)x$ corresponds to the
bandwidth tax.  Hence, the effective carrying capacity of a network,
i.e., net the bandwidth tax, can be significantly less than its raw
switching capacity; aggregate tax rates of 200--500\% are
common in existing proposals.

Reconfigurable networks seek to reduce the overall bandwidth tax rate of a given
workload by provisioning direct links between end points with the highest
demands, eliminating the tax on the largest, ``bulk'' flows
whose completion time is gated by available network capacity, rather
than propagation delay.
The time required to identify such flows~\cite{reactor:nsdi14,cthrough} and reconfigure the
network~\cite{projector,rotornet:sigcomm17}, however,
is generally orders-of-magnitude larger than the one-way delay of
even an indirect route through the network, which is the main
driver of optimal completion times for small flows.
Hence, dynamic networks face a fundamental trade-off between
amortizing the overhead of reconfiguration against the inefficiency of
sub-optimal configurations.  The upshot is existing proposals are
either unsuitable for latency sensitive traffic (which is frequently
shunted to an entirely separate network in so-called hybrid
architectures~\cite{reactor:nsdi14,rotornet:sigcomm17,mordia:sigcomm13}),
or pay substantial bandwidth tax to provide low-latency connectivity,
especially when faced with highly dynamic or unpredictable workloads.

Opera is a network architecture that minimizes the bandwidth
tax paid by bulk traffic---which makes up the vast majority of the
bytes in today's networks~\cite{dctcp,facebook:sigcomm15}---while
ensuring low-latency delivery for the (small fraction of) traffic that
cannot tolerate added delays.  Opera implements a dynamic,
circuit-switched topology that constantly reconfigures a small number
of each top-of-rack (ToR) switch's uplinks, moving through a series of
time-varying
expander graphs (without requiring runtime circuit selection algorithms
or network-wide
traffic demand collection).  Opera's ever-changing topology ensures that every
pair of end points is periodically allocated a direct link, delivering
bandwidth-efficient connectivity for bulk traffic, while indirecting
latency-sensitive traffic over the same, low-diameter network to
provide near-optimal flow completion times.

By strategically pre-configuring the assignment of rack-to-rack circuits at each
instant in time such that those circuits form an expander graph, Opera
can always forward low-latency traffic over an expander without
waiting for any circuits to be (re\nobreakdash-)configured.  Thus, on
a per-packet basis, Opera can choose to either (1) immediately send a
packet over whatever static expander is currently instantiated,
incurring a modest tax on this small fraction of traffic, or (2)
buffer the packet and wait until a direct link is established to the
ultimate destination, eliminating the bandwidth tax on the vast
majority of bytes.  Our simulation results show this trade-off results
in up to a 4$\times$ increase in throughput for shuffle workloads
compared to cost-equivalent static topologies.  Moreover, for
published, skewed datacenter workloads, Opera delivers an effective
8.4\% bandwidth tax rate, resulting in up to a 60\% increase in throughput
while maintaining equivalent flow completion times across all flow sizes.
We further validate the stability of this result across a range of
workloads, network scales, and cost factors.

 \section{Network efficiency}

The reality of datacenter networks is one of non-stop change:
developers are continuously deploying new applications and updating
existing applications, and user behavior is in a constant state of
flux.  As a result, operators cannot risk designing networks that
support only a narrow range of workloads, and instead must choose a
design that supports a wide range of workloads, applications, and user
behavior.

\subsection{Workload properties}

\begin{figure}
\centering
\includegraphics[width=0.95\columnwidth]{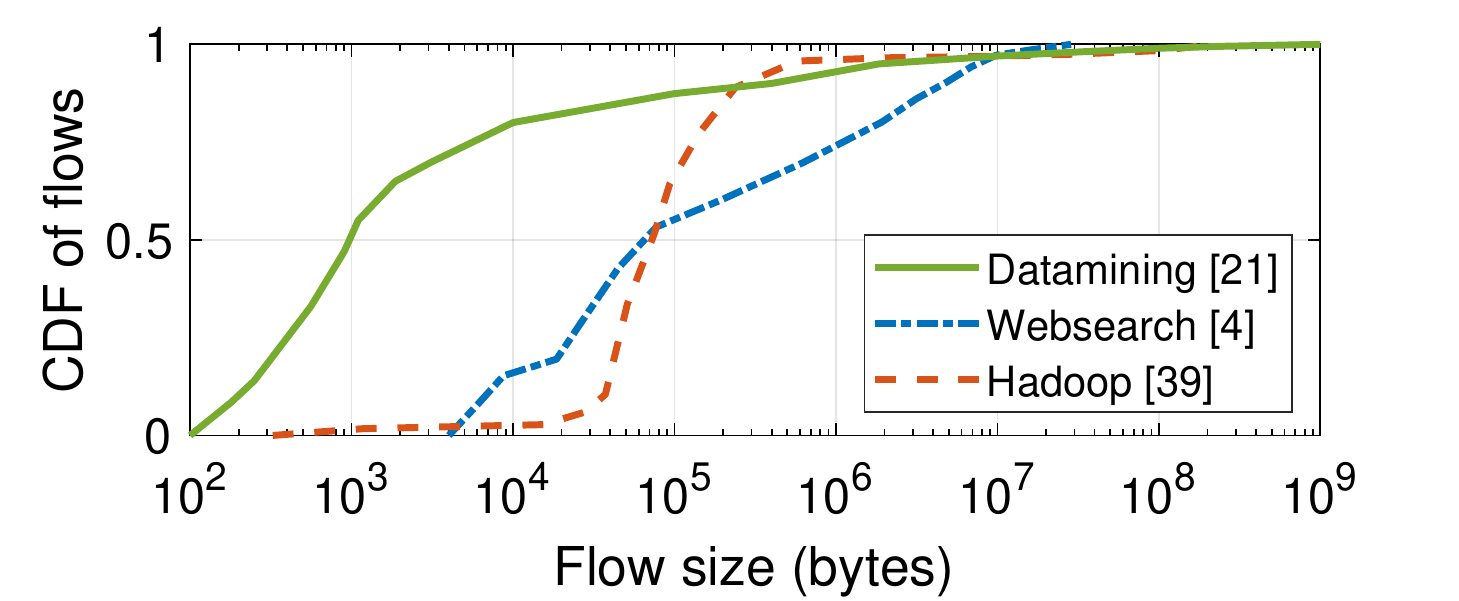}
\includegraphics[width=0.95\columnwidth]{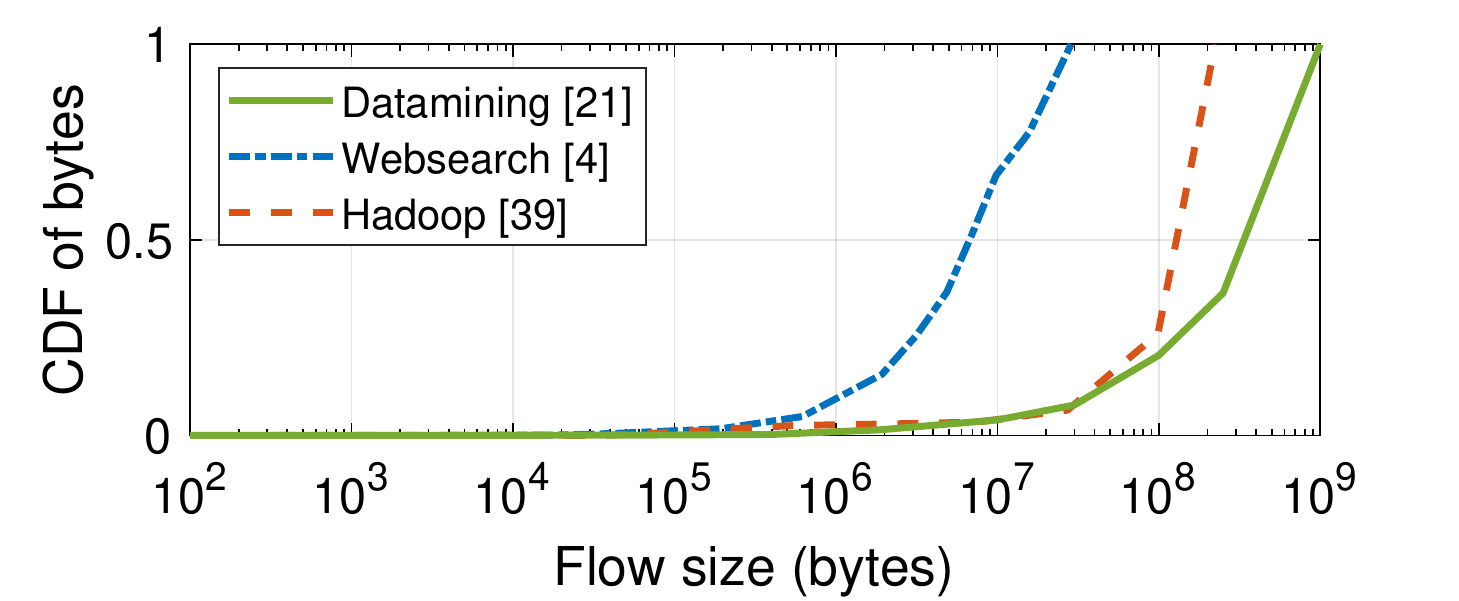}
\caption{\label{fig:flowdists} Published empirical flow-size
  distributions.}
\vskip -1em
\end{figure}

\begin{figure*}[t]
\centering
\includegraphics[width=.6\textwidth]{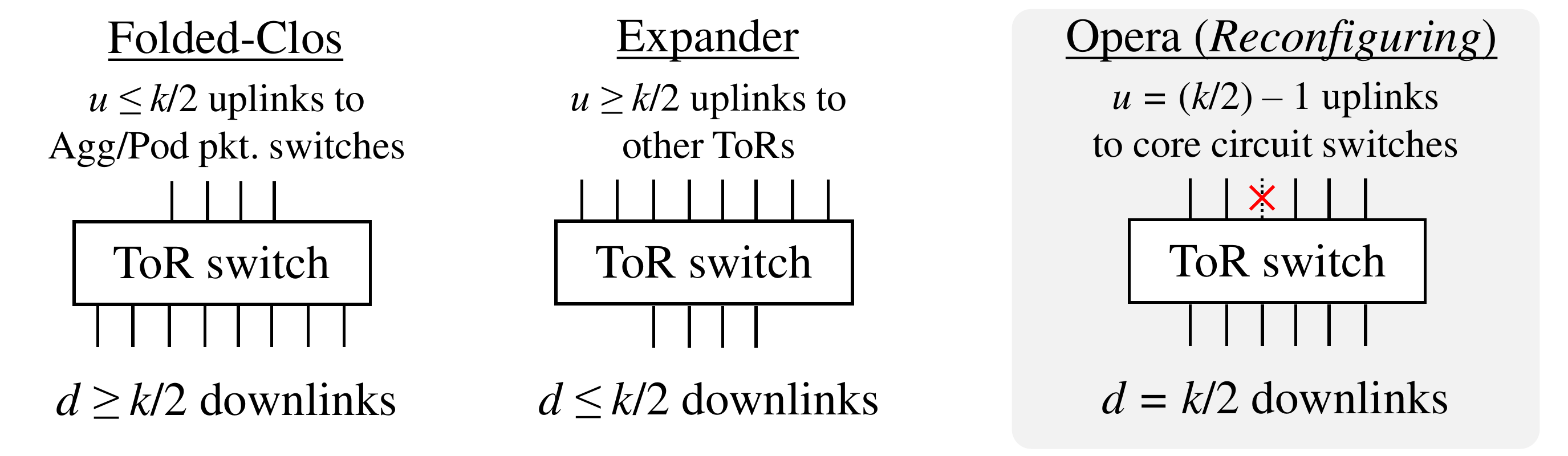}
\caption{\label{fig:tors} Oversubscribed folded-Clos networks
allocate fewer uplinks than downlinks, and static expander-graph-based networks typically
allocate more upward ports than downward ports.  In Opera, the ToR
switch is provisioned 1:1. When the circuit switch is reconfiguring, the
associated ToR port cannot carry traffic through that uplink.}
\vskip -1em
\end{figure*}

One saving grace of the need to service a wide range of workloads is
the likelihood that there will, in fact, be a spectrum of needs in
practice.  A concrete example is the distribution of flow sizes,
which is known to be highly skewed in today's networks:
Figure~\ref{fig:flowdists} shows data published by
Microsoft~\cite{dctcp,vl2} (Websearch and Datamining) and
Facebook~\cite{facebook:sigcomm15} (Hadoop) depicting the
distributions of traffic according to individual flows (top) and total
number of transmitted bytes (bottom) that we consider in this paper.
The vast majority of bytes are in bulk flows,
not the short, latency-sensitive ones, suggesting that to make the
most out of available capacity, an ideal network must seek to minimize
the bandwidth tax paid on bulk traffic while not substantially
impacting the propagation delay experienced by short flows.

While there are myriad ways to measure a network's suitability for a
given workload, flow completion time (FCT) is frequently offered as a
useful figure of merit~\cite{fctccr} due to its applicability across a
wide range of workloads.  The flow completion time of small flows is
constrained by the underlying network's propagation delay.
Thus, lowering
the network diameter and/or reducing queuing
reduces the FCT for this type of traffic.  On the other hand, the FCT
of bulk traffic is governed by the available capacity along a flow's
path.

Because the FCT of short flows is dictated by propagation delay, such
traffic is commonly referred to as ``latency-sensitive'' or,
equivalently, ``low-latency''.  (While applications may be equally
sensitive to the completion time of larger flows, their FCT is
dominated by available bandwidth.)  In today's networks, flows are
classified into these categories either explicitly (e.g., by
application type, port number, or sender-based rules), or implicitly
(e.g., by remaining flow size for shortest-remaining-time-first (SRTF)
scheduling).  Opera is agnostic to the manner in which traffic is
classified; for our purposes latency-sensitive and short flows are
synonymous.  Because latency-sensitive traffic's overall impact on
network capacity is negligible in today's workloads, it suffices to
use priority queuing to ensure short flows receive unimpeded service
while allowing bulk traffic to consume any remaining
capacity~\cite{pias,queuejump}.  The challenge is to simultaneously
provide high-capacity paths while maintaining a short path length.

\subsection{The ``big switch'' abstraction}

If cost (and practicality) were no object, a perfect network would
consist of one large, non-blocking switch that connects all the end
points.  It is precisely such a ``big switch'' illusion that scale-out
packet-switched network fabrics based on folded-Clos
topologies~\cite{dcswitch:sigcomm08,vl2,portland} were designed to provide.
These topologies rely on multiple stages of packet switches
interconnected with shuffle networks.  The abundance of packet
switches at each stage and surfeit of links between them ensures that
there is sufficient capacity to support any mixture of (admissible)
inter-server communication.
Proposals such as Hedera~\cite{hedera:nsdi10}, pHost~\cite{phost},
HULL~\cite{hull}, NDP~\cite{ndp}, PIAS~\cite{pias}, and Homa~\cite{homa} introduce flow
scheduling techniques that assign traffic to well-chosen paths to
maximize throughput while minimizing in-network queuing when servicing
a mixture of bulk and low-latency traffic.

\subsection{Reduced capacity networks}

While full-bandwidth ``big switch'' network designs are ideal in the
sense that they provide operators with the maximum flexibility to
deploy services, schedule jobs, and disaggregate storage and compute,
they are impractical to construct at scale.  Indeed,
published reports confirm the largest datacenter networks in
existence, while based upon folded-Clos topologies, are not fully
provisioned~\cite{fb-fabric,jupiter}.  Moreover,
some have observed that packet-switching
technology may not be able to keep up as link rates surpass 400 Gb/s,
so it is unclear how much longer the ``big switch'' abstraction will
even be feasible~\cite{ptree:hotnets16}.
Hence, researchers and practitioners alike have considered numerous
ways to under-provision or ``over-subscribe'' network topologies.

One way to view over-subscription in a rack-based datacenter is to
consider how each individual ToR switch is provisioned.
Consider a scenario in which servers in a cluster or datacenter
are organized into racks, each with a $k$-radix ToR packet switch
that connects it to the remainder of the network.  We say that a ToR
with $d$ connected servers has $d$ ``downward'' facing ports.  A ToR
with $u$ ports connected to the rest of the network has $u$ ``upward''
facing ports, or uplinks. (In a fully populated ToR, $d+u=k$.)  In this
context, we now overview existing proposals of how to interconnect
such racks.

\paragraph{Over-subscribed Fat Trees:}
As shown in the left-most portion of Figure~\ref{fig:tors}, designers
can build $M$:1 over-subscribed folded-Clos networks in which the
network can deliver only $(1/M=u/d)$ the bandwidth of a
fully-provisioned design.  Common values of $(d:u)$ are between 3:1
and 5:1~\cite{jupiter}.  The cost and bandwidth delivered in
folded-Clos networks scale almost linearly according to the
over-subscription factor, and so decreasing overall cost necessitates
decreasing the maximum network throughput---and vice versa.  Routing
remains direct, however, so over-subscription does not introduce a
bandwidth tax; rather, it severely reduces the available network
capacity between end points in different racks.  As a result,
application frameworks such as MapReduce~\cite{mapreduce} and
Hadoop~\cite{hadoop} schedule jobs with locality in mind in an effort
to keep traffic contained with a rack.

\paragraph{Expander topologies:} To address the limited cross-network bandwidth available in
over-subscribed Fat Trees, researchers have proposed alternative
reduced-capacity network topologies based on expander graphs.  In
these proposals, the $u$ uplinks from each ToR are directly connected
to other ToRs, either randomly~\cite{jellyfish} or
deterministically~\cite{expander1,expander2,xpander}, reducing the
number of switches and inter-switch links internal to the network
itself.  Expander-graph-based network topologies are sparse graphs with the
property that there are many potential short paths from a given source
to a particular destination.

Because there are no in-network
switches,
packets must ``hop''
between ToRs a number of times to reach their ultimate destination,
resulting in a bandwidth tax.
An expander graph with an average ToR-to-ToR hop count of $L_{Avg}$
pays an overall bandwidth tax rate of $(L_{Avg}-1)\times$ in expectation
because individual packets must indirect across a number of 
in-network links.  The average path lengths for large networks can be
in the range of 4--5 hops, resulting in a bandwidth tax rate of 300--400\%.
Moreover, a recent proposal~\cite{expander2} employs Valiant load
balancing (VLB)---which imposes an additional level of explicit
indirection---to address skewed traffic demands, doubling the
bandwidth tax in some circumstances.  One way that expanders
counter-act their high bandwidth tax rate is by over-provisioning:
ToRs in expander topologies typically have more upward-facing ports
than down ($u > d$, as shown in the center of
Figure~\ref{fig:tors})---and, hence, far more upward-facing ports than
over-subscribed Fat Trees---which provides more in-network capacity.
Said another way, the impact of the bandwidth tax is reduced by a
factor of $u/d$.

\paragraph{Reconfigurable topologies:} In an effort to reduce the bandwidth tax, other proposals rely on some form of reconfigurable link technology,
including RF~\cite{flyways,3DBeam}, free-space
optical~\cite{projector,firefly}, and circuit
switching~\cite{helios:sigcomm10,reactor:nsdi14,mordia:sigcomm13,shoal,cthrough}.
Most reconfigurable topologies dynamically establish end-to-end paths
within the network core in response to traffic demand, although
RotorNet~\cite{rotornet:sigcomm17} employs a fixed, deterministic
schedule.  In either case, these networks establish and tear down
physical-layer links over time.  When the topology can be matched to
the demand---and setting aside latency concerns---traffic can be
delivered from source to destination in a single hop, avoiding any
bandwidth tax.  In some cases, similar to expander-based topologies,
they employ 2-hop VLB~\cite{rotornet:sigcomm17,shoal},
resulting in a 100\% bandwidth tax rate.

A fundamental limitation of any reconfigurable topology, however, is
that during the time a link/beam/circuit (for simplicity we will use
the latter term in the remainder of the paper) is being provisioned,
it cannot convey data.
Moreover, most proposals
do not provision links between all sources
and destinations at all times, meaning that traffic may incur
significant delay as it waits for the appropriate circuit to be
provisioned.  For existing proposals, this end-to-end delay is on
the order of 10--100s of milliseconds.
Hence, previous proposals for reconfigurable network topologies
rely on a distinct, generally packet-switched, network to service
latency-sensitive traffic.
The requirement for a separate network built out of a distinct technology
is a significant practical limitation and source of cost and power
consumption.

 \section{Design}
\label{sec:design}

We start with an overview of our design before working through an
example.  We then proceed to describe how we construct the topology of
a given network, how routes are chosen, how
the network moves through its fixed set of configurations, and address
practical considerations like cabling complexity, switching speeds,
and fault tolerance.

\subsection{Overview}

Opera is structured as a two-tier leaf-spine topology, with
packet-switched ToRs interconnected by
reconfigurable circuit switches as shown in Figure~\ref{fig:topo}.
Opera's design is based around two fundamental starting blocks that
follow directly from the requirements for small network diameter and
low bandwidth tax.

\paragraph{Expansion for short paths:} Because the flow completion time
of short, latency-sensitive flows is gated by end-to-end delay, we
seek a topology with the lowest possible expected path
length. Expander-based topologies are known to be ideal~\cite{expander1}.
Expanders also have
good fault-tolerance properties; if switches or links fail, there are
likely to be alternative paths that remain.  Thus,
to efficiently support low-latency traffic, we require a topology with
good expansion properties at all times.

\paragraph{Reconfigurability to avoid the bandwidth tax:} 
A fully-connected graph (i.e. full mesh) could avoid a bandwidth tax
entirely, but is infeasible to construct at scale. Rather than
providing a full mesh in space, reconfigurable circuit switches offer
the ability to establish, over time, direct one-hop paths between
every rack pair using a relatively small number of links.  Because
bulk flows can generally amortize modest reconfiguration overheads if
they result in increased throughput, we incorporate reconfigurability
into our design to minimize the bandwidth tax on bulk traffic.

Opera combines the elements of expansion and reconfigurability to
efficiently (and simultaneously) serve both low-latency and bulk
traffic with low flow completion time. Similar to
RotorNet~\cite{rotornet:sigcomm17}, our design incorporates
reconfigurable circuit switches that cyclically set up and tear down
direct connections between ToRs, such that after a ``cycle time'' of
connectivity, every ToR has been connected to every other ToR.  We
take advantage of ToR-uplink parallelism to stagger the
reconfigurations of multiple circuit switches, allowing ``always-on''
multi-hop connectivity between all ToR pairs.

Critically, the combination of circuits at any time forms an expander graph.
Thus, during a single cycle, every packet has a choice between waiting for a
bandwidth-tax-avoiding direct connection, or being immediately sent over a multi-hop path through the time-varying expander.
The end result
is a single fabric that supports bulk and low-latency traffic
as opposed to two separate networks used in hybrid approaches.
As we will show, Opera does not require any runtime
selection of circuit assignments or
system-wide collection of traffic demands, vastly simplifying its
control plane relative to dynamic approaches such a ProjecToR~\cite{projector}
and Mordia~\cite{mordia:sigcomm13}.

\subsubsection{Eliminating reconfiguration disruptions}
\label{sec:design:offset}

\begin{figure}
\centering
\begin{subfigure}[b]{.49\columnwidth}
\includegraphics[width=\linewidth]{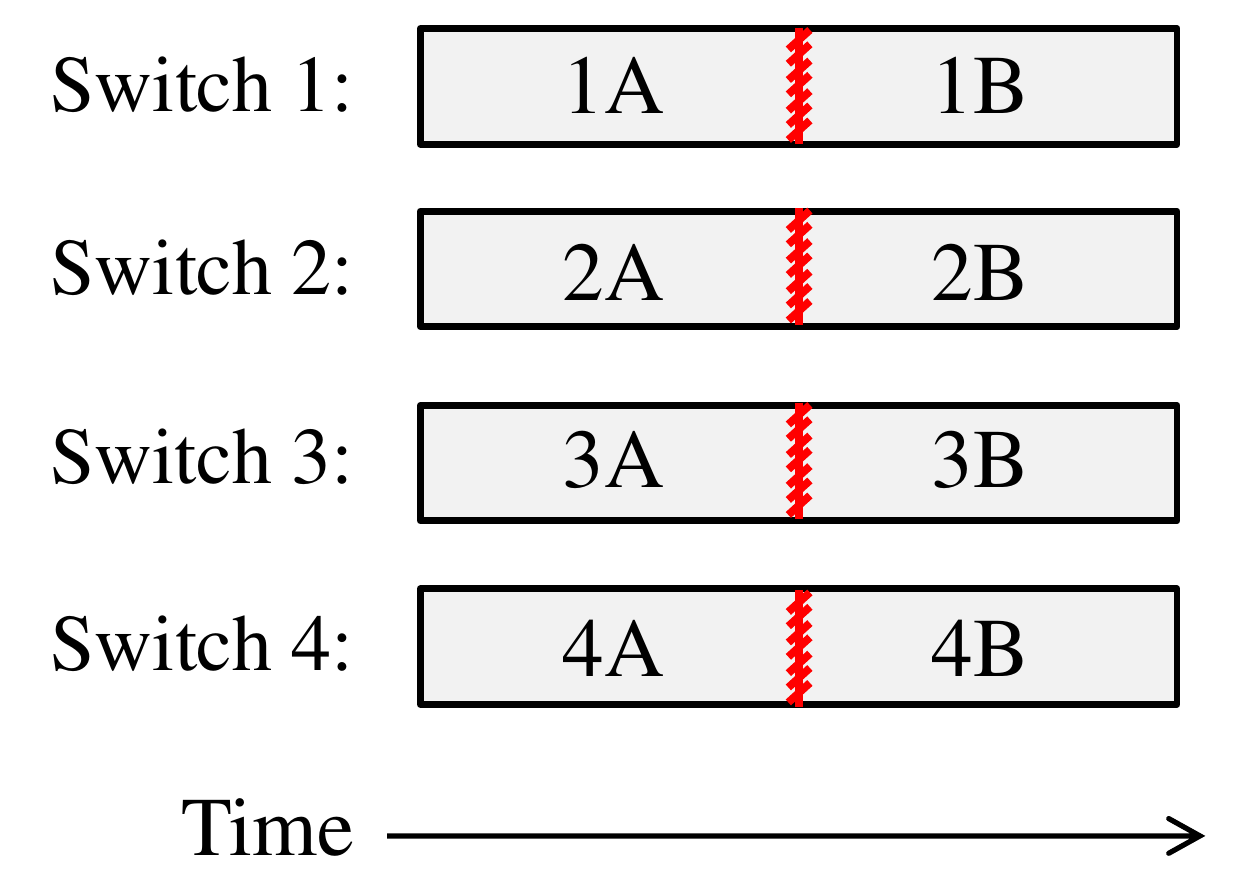}
\caption{\label{fig:offset_a} Simultaneous reconfig.}
\end{subfigure}
\begin{subfigure}[b]{0.49\columnwidth}
\includegraphics[width=\linewidth]{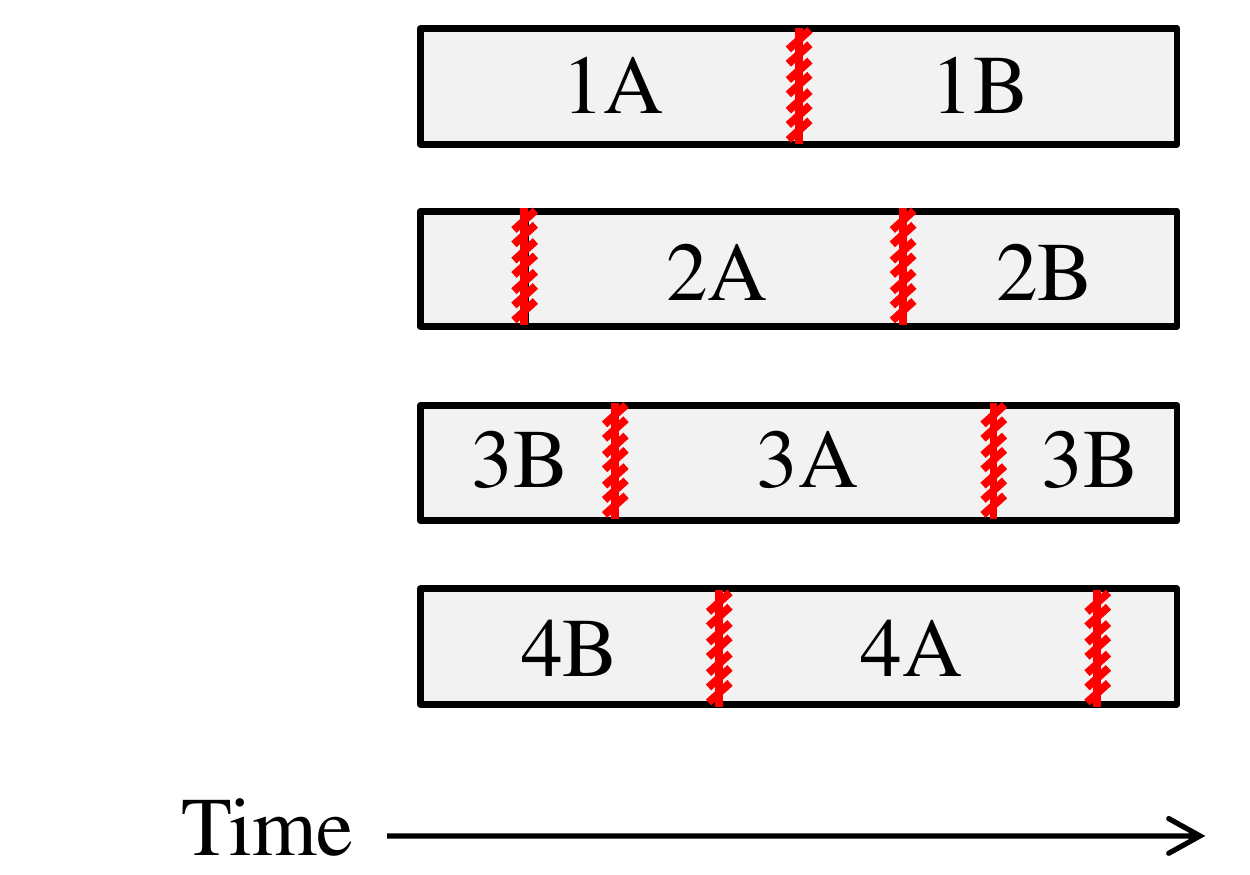}
\caption{\label{fig:offset_b} Offset reconfiguration}
\end{subfigure}
\caption{\label{fig:offset} Reconfiguring all switches in unison (a) leads
to periodic disruptions; staggered reconfigurations (b) ensure some paths 
are always available.}
\vskip -1em
\end{figure}

Circuit switches impose a technology-dependent reconfiguration
delay, necessitating that flows be re-routed before reconfiguration.
Even in a network with multiple circuit switches, if
all switches reconfigure simultaneously (Figure~\ref{fig:offset_a}),
the global disruption in connectivity requires routes to
reconverge. For today's switching technologies, this would
lead to traffic
delays that could severely impact the flow completion time of short,
latency-sensitive flows.
To avoid this scenario and allow for low-latency packet delivery,
Opera offsets the reconfigurations of circuit switches. For
example, in the case of small topologies with few switches, at most
one switch may be reconfiguring at a time (Figure~\ref{fig:offset_b}),
allowing flows traversing a circuit with an impending reconfiguration
to be migrated to other circuits that will remain active during that
time period (for large-scale networks with many circuit switches, it is
advantageous to reconfigure more than one switch at a time as described
in Appendix~\ref{app:cycle}). As a result, while Opera is in
near-constant flux, changes are incremental and connectivity is
continuous across time.

\subsubsection{Ensuring good expansion}
\label{sec:design:topology}

While offsetting reconfigurations guarantees continuous connectivity,
it does not, by itself, guarantee complete connectivity. Opera must
simultaneously ensure that (1) multi-hop paths exist between all racks
at every point in time to support low-latency traffic, and (2) direct
paths are provisioned between every rack-pair over a fixed period of
time to support bulk traffic with low bandwidth tax.
We guarantee both by implementing a (time-varying) expander graph
across the set of circuit switches.

In Opera, each of a ToR's $u$ uplinks is connected to a (rotor)
circuit switch~\cite{jlt} that, at any point in time,
implements a (pre-determined) random permutation between input and
output ports (i.e., a ``matching'').  The inter-ToR network topology
is then the union of $u$ random matchings, which, for $u\geq3$, results
in an expander graph with high probability~\cite{alon:expanders}.
Moreover, even if a switch is reconfiguring, there are still $u-1$
active matchings, meaning that if $u\geq4$, the network will still be an
expander with high probability, no matter which switch is
reconfiguring.  In Opera, $u=k/2$, and $k$ is on the order of 10s of
ports for today's packet switches.

\begin{figure}
  \centering
  \vskip -0.5em
\includegraphics[width=0.85\columnwidth]{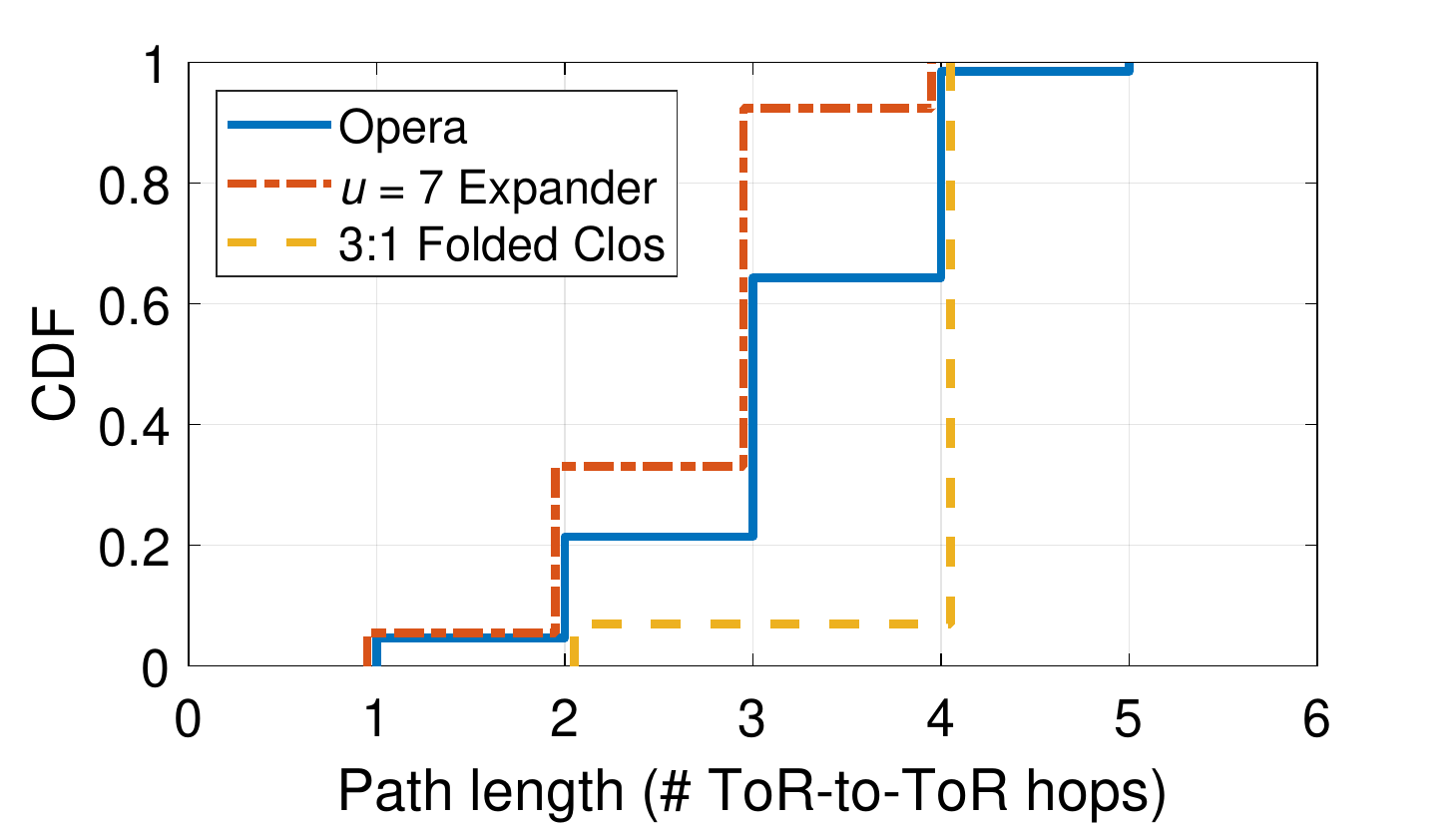}
\caption{\label{fig:hops2} CDF of path lengths for equal-cost 648-host
  Opera, 650-host $u=7$ expander, and 648-host 3:1 folded-Clos
  networks.
    (CDFs staggered slightly for clarity.)}
\vskip -1em
\end{figure}

Figure~\ref{fig:hops2} shows the distribution of path lengths in one
example 648-host network considered in our evaluation, where $u=6$.
Opera's path lengths are almost always substantially shorter than
those in a Fat Tree that connects the same number of hosts, and only
marginally longer than an expander with $u=7$ which we argue later has
similar cost, but performs poorly for certain workloads.  Clearly,
ensuring good expansion alone is not an issue with modest switch
radices.  However, Opera must also directly connect each rack pair
over time.  We achieve this by having each switch cycle through a set
of matchings; we minimize the total number of matchings by
constructing a disjoint set.

\begin{figure*}
  \centering
\begin{subfigure}[b]{.46\textwidth}
\includegraphics[width=\linewidth]{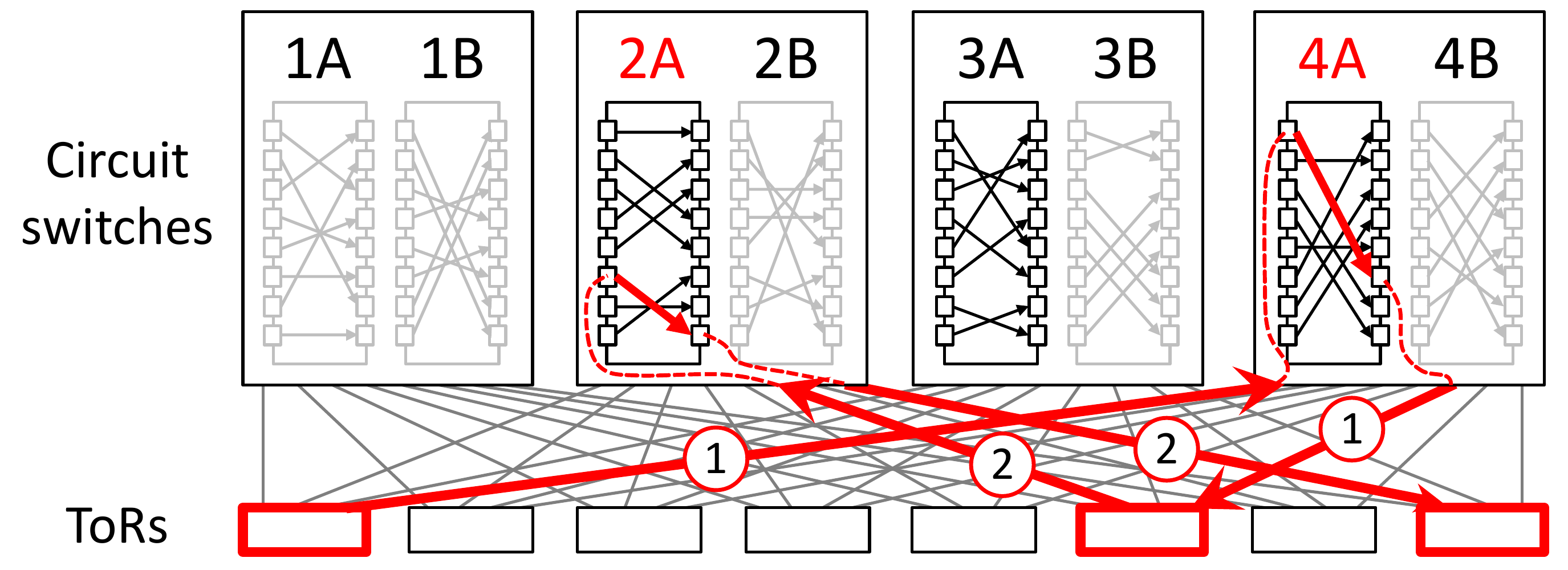}
\caption{\label{fig:topo_a} Indirect path}
\end{subfigure}
\hspace{0.2in}
\begin{subfigure}[b]{.40\textwidth}
\includegraphics[width=\linewidth]{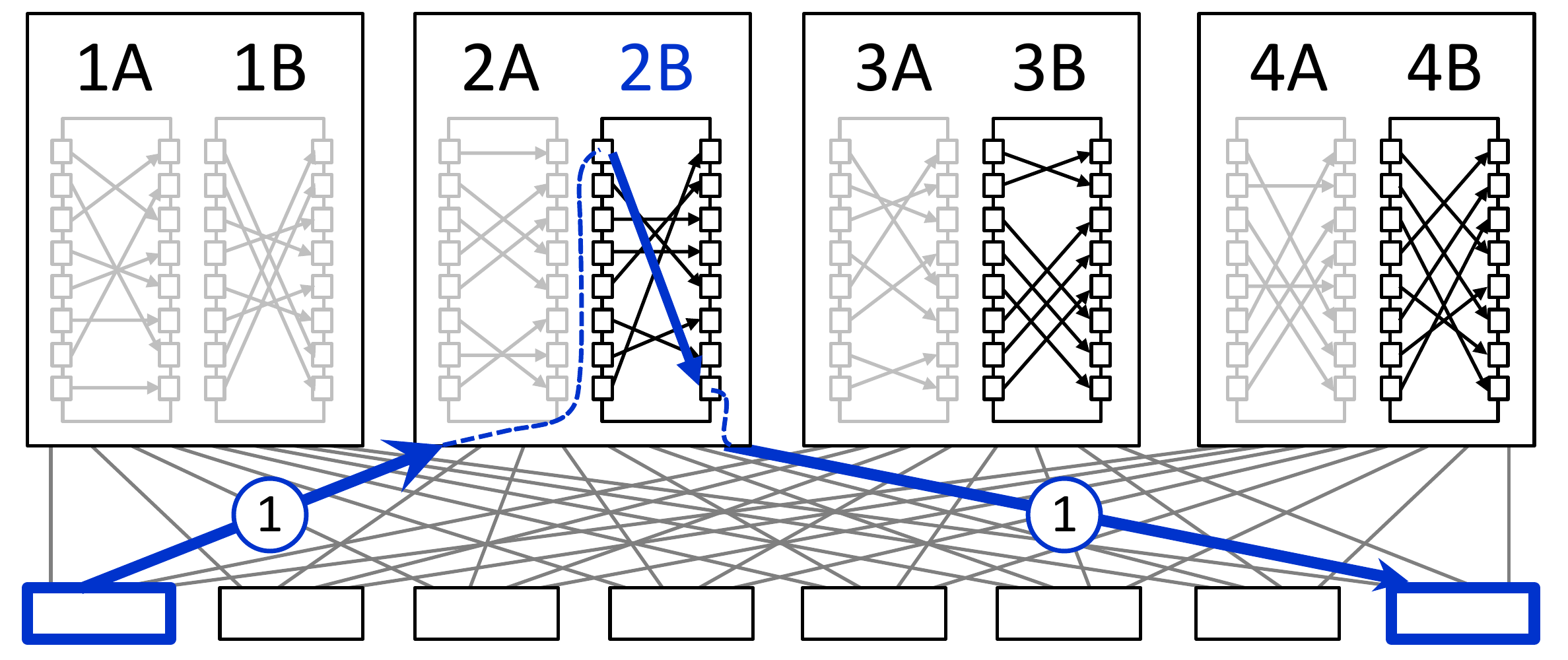}
\caption{\label{fig:topo_b} Direct path}
\end{subfigure}
\caption{ \label{fig:topo} An Opera topology with eight ToR switches
  and four rotor circuit switches (from
  RotorNet~\cite{rotornet:sigcomm17}).  Two different paths from rack
  1 to rack 8 are highlighted: (a) a two-hop path in red, and (b) a
  one-hop path in blue.  Each direct inter-rack connection is
  implemented only once per configuration, while multi-hop paths are
  available between each rack-pair at all times.}
\vskip -1em
\end{figure*}

\subsection{Example}

Figure~\ref{fig:topo} depicts a small-scale Opera network.  Each of
the eight ToRs has four uplinks to four different circuit switches
(with one potentially down due to reconfiguration at any particular
moment).
By forwarding traffic through those ToRs, they can reach any ToRs
to which they, in turn, are connected.
Each circuit switch has two matchings, labeled $A$ and $B$ (note that
all matchings are disjoint from one another). In this example
topology, any ToR-pair can communicate by utilizing any set of three
matchings, meaning complete connectivity is maintained regardless of
which matchings happen to be implemented by the switches at a given
time.  Figure~\ref{fig:topo} depicts two network-wide configurations.
In Figure~\ref{fig:topo_a} switches 2--4 are implementing matching
$A$, and in Figure~\ref{fig:topo_b}, switches 2--4 implement matching
$B$. In both cases switch 1 is unavailable due to reconfiguration.

In this example, racks 1 and 8 are directly connected by the
configuration shown in Figure~\ref{fig:topo_b}, and so the lowest
bandwidth-tax way to send bulk data from 1 to 8 would be to wait until
matching $B$ is instantiated in switch 2, and then to send the data
through that circuit; such traffic would arrive at ToR 8 in a single
hop.  On the other hand, low-latency traffic from ToR 1 to ToR 8
can be sent immediately, e.g. during the configuration shown in
Figure~\ref{fig:topo_a}, and simply take a longer path to get to ToR
8.  The traffic would hop from ToR 1 to ToR 6 (via switch 4), then to
ToR 8 (via switch 2), and incur a 100\% bandwidth tax.  Although not
highlighted in the figure, similar alternatives exist for
all rack pairs.

\subsection{Topology generation}

The algorithm to generate an Opera
topology is as follows. First, we randomly factor a complete graph
(i.e. $N \times N$ all-ones matrix) into $N$ disjoint (and symmetric)
matchings. Because this factorization can be computationally expensive
for large networks, we employ graph lifting to generate large
factorizations from smaller ones. Next, we randomly assign the $N$
matchings to circuit switches, so that each switch has $N/u$ matchings
assigned to it. Finally, we randomly choose the order in which each
switch cycles through its matchings. These choices are fixed at
design time, before the network is put into operation; there is no
topology computation during network operation.

Because our construction approach is random, it is possible (although
unlikely) that a specific Opera topology realization will not have good
expander properties at all points across time. For example, the combination
of matchings in a given set of $u-1$ switches at a particular time may
not constitute an expander. In this case, it would be trivial to generate
and test additional realizations at design time until a solution with
good properties is found. This was not necessary in our experience, as
the first iteration of the algorithm always produced a topology with
near-optimal properties.  We discuss the properties
of these graphs in detail in Appendix~\ref{app:graph}.

\subsection{Forwarding}

We are now left to decide how to best serve a given flow or packet:
(1) send it immediately over multi-hop expander paths and pay the
bandwidth tax (we refer to these as ``indirect'' paths), or (2)
delay transmission and send it over one-hop paths to avoid the
bandwidth tax (we refer to these as ``direct'' paths).  For skewed
traffic patterns that can tolerate delay, two-hop paths based on
Valiant load balancing can be used as well.
Our baseline approach is to decide based on the flow size.
Since the delay in waiting for a direct path can be an entire cycle
time, we only let flows that are long enough to amortize that delay
use direct paths, and place all other traffic on indirect paths.
However, we can do even better if we know something about
application behavior. Consider an all-to-all shuffle operation, where
a large number of hosts simultaneously need to exchange a small amount
of data with one another. Although each flow is small, there will be
significant contention, extending the flow completion time of
these flows. Minimizing bandwidth tax is critical in these situations.
With application-based tagging, Opera can route such traffic over
direct paths.

\subsection{Synchronization}
\label{sec:sync}

Opera employs reconfigurable circuit switches, and so its design requires a
certain level of synchronization within the system to operate correctly.  In
particular, there are three synchronization requirements that must be met: (1)
ToR switches must know when core circuit switches are reconfiguring, (2) ToR
switches must update their forwarding tables in sync with the changing core
circuits, and (3) end hosts must send bulk traffic to their local ToR during
times when the ToR is directly connected to the destination (to prevent
excessive queueing in the ToR).
In the first case, since each ToR's uplink is connected directly to
one of the circuit switches, the ToR can monitor the signal strength
of the transceiver attached to that link to re-synchronize with the
circuit switch.  Alternatively, the ToR could rely on IEEE 1588 (PTP),
which can synchronize switches to within $\pm$1 $\upmu$s~\cite{ptp}.
For low-latency traffic, end hosts
simply transmit packets immediately, without any coordination or
synchronization.  For bulk traffic, end hosts transmit
when polled by their attached ToR.
To evaluate the practicality of this synchronization approach, we built
a small-scale prototype based on a programmable P4 switch, described
in Section~\ref{sec:discussion:prototype}.

Opera can tolerate arbitrary bounds on
(de\nobreakdash-)synchronization by introducing ``guard bands'' around
each configuration, in which no data is sent to ensure the network is
configured as expected when transmissions do occur. In our design, each
$\upmu$s of guard time contributes a 1\% relative reduction in low-latency
capacity and a 0.2\% reduction for bulk traffic. In practice, if
any component becomes de-synchronized beyond the guard-band tolerance,
it can simply be declared failed (see Section~\ref{s:ft}).

\subsection{Practical considerations}

While Opera's design draws its power from graph-theoretic
underpinnings, it is also practical to deploy.
Here, we consider two important constraints on real-world networks.

\subsubsection{Cabling and switch complexity}

Today's datacenter networks are based on folded-Clos topologies which
use perfect-shuffle cabling patterns between tiers of switches.
While proposals for static expander graphs alter that wiring
pattern~\cite{jellyfish} leading to concerns about cabling complexity,
Opera does not. In Opera, the interconnection complexity is contained
within the circuit switches themselves, while the inter-switch cabling
remains the familiar perfect shuffle.
In principle, Opera can be implemented with a variety of electronic or
optical circuit switch technologies. We focus on optical switching for
our analysis due to its cost and data-rate transparency
benefits. Further, because each circuit switch in Opera must only
implement $N/u$ matchings (rather than $O(N!)$), Opera can make use of
optical switches with limited configurability such as those proposed
in RotorNet~\cite{rotornet:sigcomm17}), which have been demonstrated
to scale better than optical crossbar
switches~\cite{Ford:94,jlt}.

\subsubsection{Fault tolerance}
\label{s:ft}

Opera recovers from link, ToR, and circuit switch failures using common
routing protocol practices: ToRs use a ``hello'' protocol initiated at the
beginning of each new matching to both detect
failures and share failure information with other ToRs. Upon receiving
information of a new failure, a ToR recomputes and updates its routing
tables to route around failed components.
We take advantage of Opera's cyclic connectivity to detect and
communicate failures: each time a new circuit is configured, the ToR
CPUs on each end of the link exchange a short sequence of hello
messages (which also contain information of new failures, if
applicable). If no hello messages are received within a configurable
amount of time, the ToR marks the link in question as bad. Because all
ToR-pair connections are established every cycle, any ToR that remains
connected to the network will learn of any failure event within at
most two cycles (1--10ms).

 \section{Implementation}

Here, we describe the implementation details of Opera. To ground our
discussion, we refer to an example 108-rack, 648-host topology based on $k=12$
(we generalize this analysis in our evaluation).

\subsection{Defining bulk and low-latency traffic}

In Opera, traffic is defined as low-latency if it cannot wait until a direct
bandwidth-efficient path becomes available.  Thus the division between
low-latency and bulk traffic depends on the rate at which Opera's circuit
switches cycle through direct matchings.  The faster Opera steps through these
matchings, the lower the overhead for sending traffic on direct paths, and thus
the larger the fraction of traffic that can utilize these paths. Two factors
impact cycle speed: circuit amortization and end-to-end delay.

\paragraph{Circuit amortization:}
The rate at which a circuit switch can change matchings is technology
dependent.  State-of-the-art optical switches with the port count and
insertion loss properties needed for practical datacenter deployment
have reconfiguration delays on the order of
10$~\upmu$s~\cite{rotornet:sigcomm17,mordia:sigcomm13,projector}.  A
90\% amortization of this delay would limit circuit reconfiguration
events to every 100~$\upmu$s. For large networks employing parallel
circuit switches, approximately 10--20 such matchings would be
necessary~\cite{rotornet:sigcomm17}, meaning that any flow than can
amortize a 1--2 ms increase in its FCT could take
the bandwidth-efficient direct paths (and shorter flows would take
indirect paths).

\begin{figure}[t]
\centering
\includegraphics[width=\columnwidth]{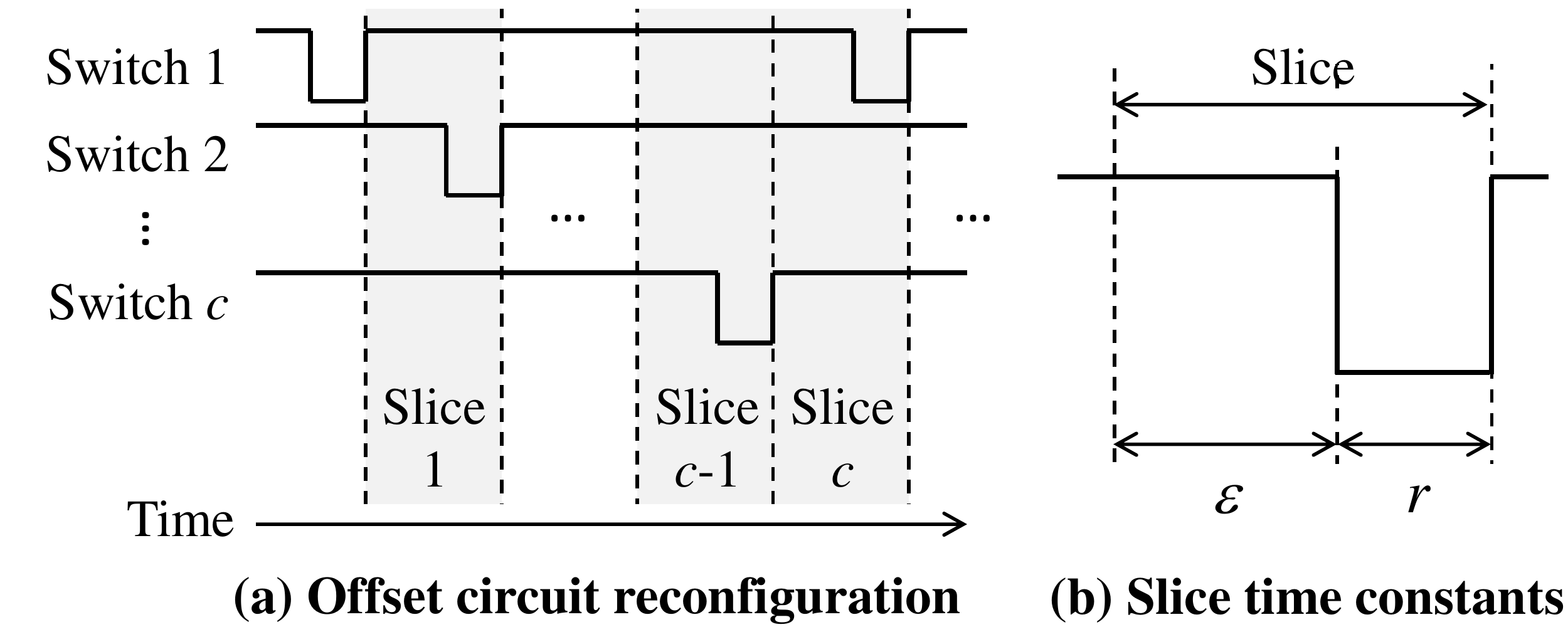}
\caption{\label{fig:offset_detail} (a) A set of $c$ circuit switches
  with offset reconfigurations forms a series of topology slices. (b)
  The time constants associated with a single slice: $\varepsilon$ is
  the worst-case end-to-end delay for a low-latency packet to traverse
  the network and $r$ is the circuit switch reconfiguration delay.}
\vskip -1em
\end{figure}

\paragraph{End-to-end delay:}
Perhaps surprisingly, a second timing constraint, end-to-end delay,
has a larger impact on cycle time.  In particular, consider a
low-latency packet that is emitted from a host NIC.  At the first ToR,
the packet is routed toward its destination, and in general, at each
hop along the way, each ToR routes the packet along an expander-graph
path.  If, during the packet's journey, the circuit topology changes,
it is possible the packet could be caught in a loop or redirected
along a sub-optimal path.  Dropping the packet immediately (and
expecting the sender to resend it) would significantly delay the flow
completion time of that flow.

Our approach, depicted in Figure~\ref{fig:offset_detail}, to avoid the
problems described above, requires that subsequent circuit
re-configurations be spaced by at least the sum of the end-to-end delay
under worst-case queuing, $\varepsilon$, and the reconfiguration
delay, $r$. We refer to this time period $\varepsilon$+$r$ as a
``topology slice''. Any packets sent during a slice are not routed
through the circuit with an impending reconfiguration during that
slice. This way, packets always have at least $\varepsilon$ time to
make it through the network before a switch re-configures.

The parameter $\varepsilon$ depends on the worst-case path length (in
hops), the queue depth, the link rate, and propagation delay.
Path length is a function of the expander, while the data rate and
propagation delay are fixed; the key driver of $\varepsilon$ is the
queue depth.
As explained in the following section, we choose a shallow queue depth
of 24 KB (8 1500-byte full packets + 187 64-byte headers). When
combined with a worst-case path length of 5 ToR-to-ToR hops
(Figure~\ref{fig:hops2}), 500-ns propagation delay per hop (100 meters
of fiber), and 10-Gb/s link speed, we set $\varepsilon$ to 90
$\upmu$s. The inter-reconfiguration period on a single switch is about
$6\varepsilon$, yielding a duty cycle of 98\% and a cycle time of
10.7 ms.  For these time constants, flows $\geq$15 MB will have a
completion time well within a factor of 2 of their ideal
(link-rate-limited) FCT. As we will show in Section~\ref{sec:eval},
depending on traffic conditions, shorter flows may benefit from direct
paths as well.

\subsection{Transport protocols}
\label{sec:transport}

Opera requires transport protocols that can (1) immediately send
low-latency traffic into the network, while (2) delaying bulk traffic
until the appropriate time.  To avoid head-of-line blocking, NICs and
ToRs each perform priority queuing.

\subsubsection{Low-latency transport}

As discussed in the previous section, minimizing the cycle time is
predicated on minimizing the queue depth for low-latency packets at
ToRs. The recently proposed NDP protocol~\cite{ndp} is a promising
choice because it achieves high throughput with very shallow
queues.  We find that 12-KB queues work well for Opera (each port has
an additional equal-sized header queue).  NDP also has other
beneficial characteristics for Opera, such as zero-RTT convergence and
no packet metadata loss to eliminate RTOs. Despite being designed for
fully-provisioned folded Clos networks, we find in simulation that NDP
works well with minimal modification in Opera, despite Opera's
continuously-varying topology. Other transports, like the recently
proposed Homa protocol~\cite{homa}, may also be a good fit for
low-latency traffic in Opera, but we leave this to future work.

\subsubsection{Bulk transport}

Opera's bulk transport protocol is relatively simple. We draw heavily
from the RotorLB protocol proposed in
RotorNet~\cite{rotornet:sigcomm17}, which buffers traffic at end hosts
until direct connections to the destination are available.  When bulk
traffic is heavily skewed, and there is necessarily spare capacity
elsewhere in the network, RotorLB automatically transitions to using
two-hop routing (i.e. Valiant load balancing) to improve
throughput. Unlike low-latency traffic, which can be sent at any time,
bulk traffic admission is coordinated with the state of the circuit
switches, as described in Section~\ref{sec:sync}. In addition to
extending RotorLB to work with offset re-configurations, we also
implemented a NACK mechanism to handle cases where large bursts of
priority-queued low-latency traffic can cause bulk traffic queued at
the ToR to be delayed beyond the transmission window and dropped at
the ToR.  Retransmitting a small-to-moderate number of packets does
not significantly affect the FCT of bulk traffic.

\subsection{Packet forwarding}

Opera relies on ToR switches to route packets along direct or
multi-hop paths depending on the requested network service model.  We
implement this routing functionality using the P4 programming
language.  Each ToR switch has an in-built register that represents
the current network configuration, updated either in-band or via PTP.
When a packet arrives at the first ToR switch, it annotates the packet's
metadata with the value of the
configuration register.  What happens next, and at subsequent ToR switches,
depends on the value of the
DSCP field.  If that field indicates a low-latency packet, then the
switch consults a low-latency table to determine the next hop along
the expander path for the current configuration, and then forwards the
packet out that port.  If the field indicates bulk traffic, then the
switch consults a bulk traffic table which indicates which circuit
switch---if any---provides a direction connection, and the packet is forwarded
to that port.
We measure the amount of
in-switch memory required to implement this program for various
datacenter sizes in Section~\ref{sec:discussion:feasibility}.
 \section{Evaluation}
\label{sec:eval}

\begin{figure*}[t]
\centering
\begin{subfigure}[b]{.24\textwidth}
\includegraphics[width=\linewidth]{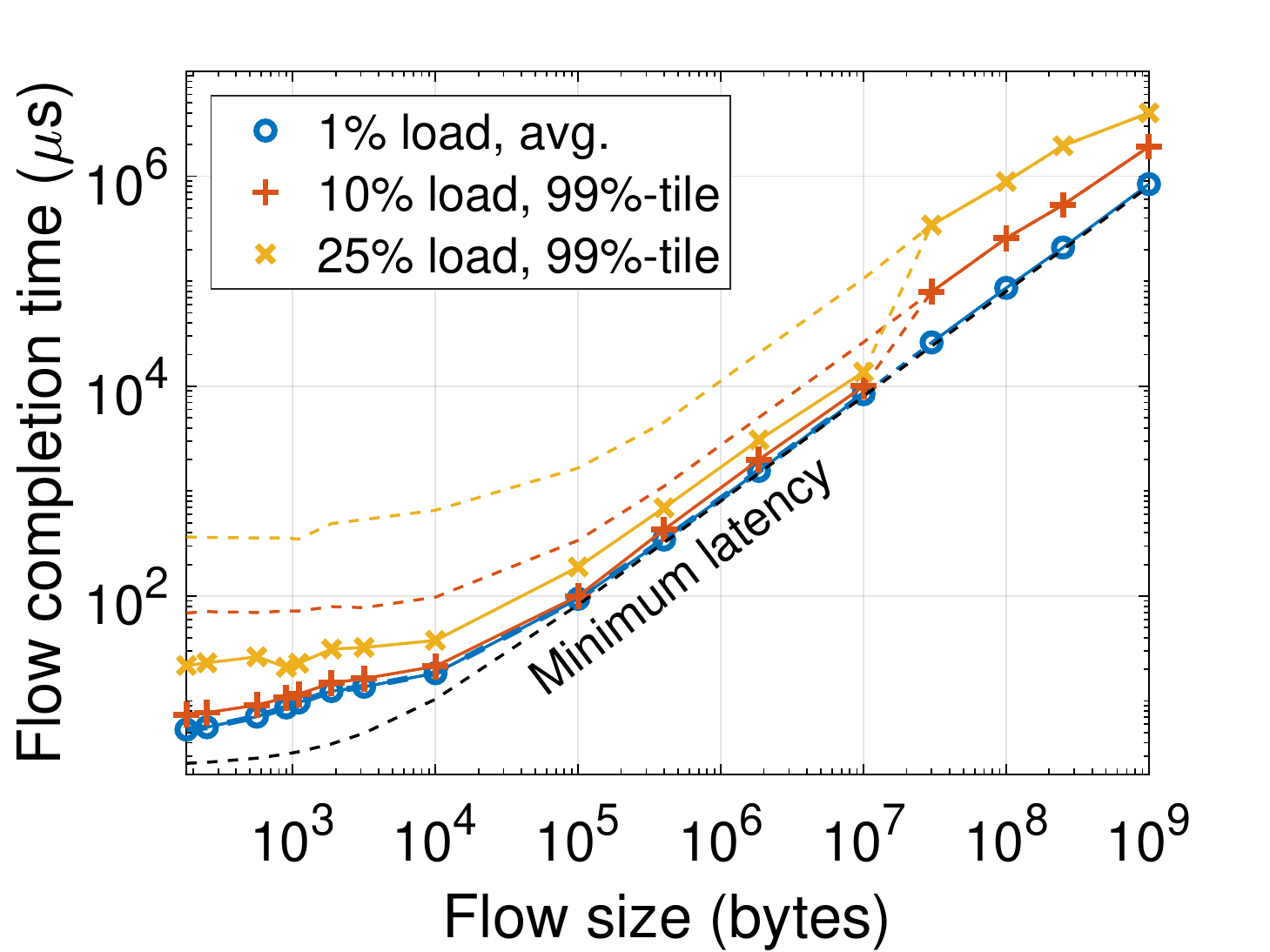}
\caption{\label{fig:fcts_3to1_1} 3:1 folded Clos}
\end{subfigure}
\begin{subfigure}[b]{0.24\textwidth}
\includegraphics[width=\linewidth]{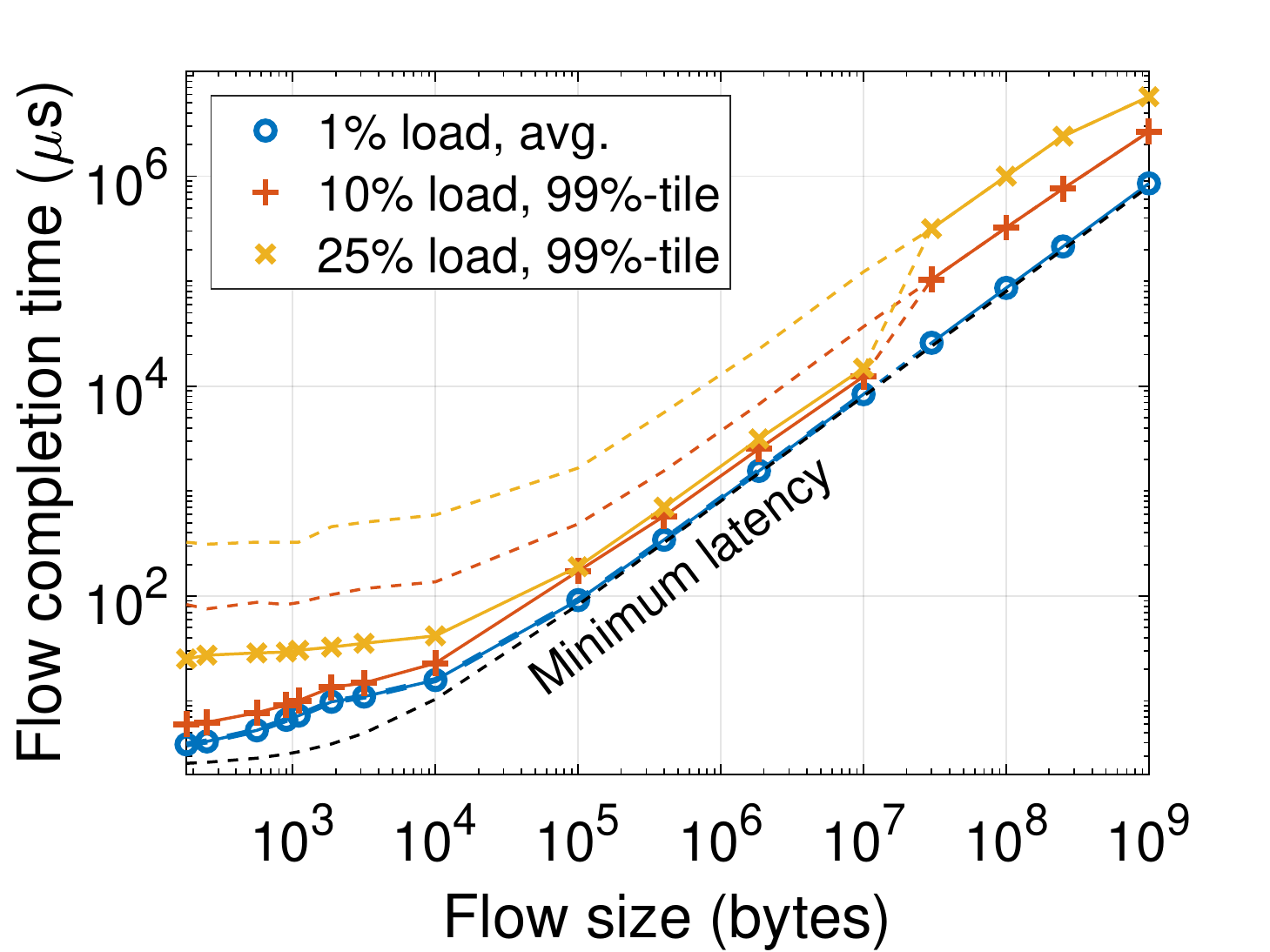}
\caption{\label{fig:fcts_u7_1} $u=7$ expander}
\end{subfigure}
\begin{subfigure}[b]{0.24\textwidth}
\includegraphics[width=\linewidth]{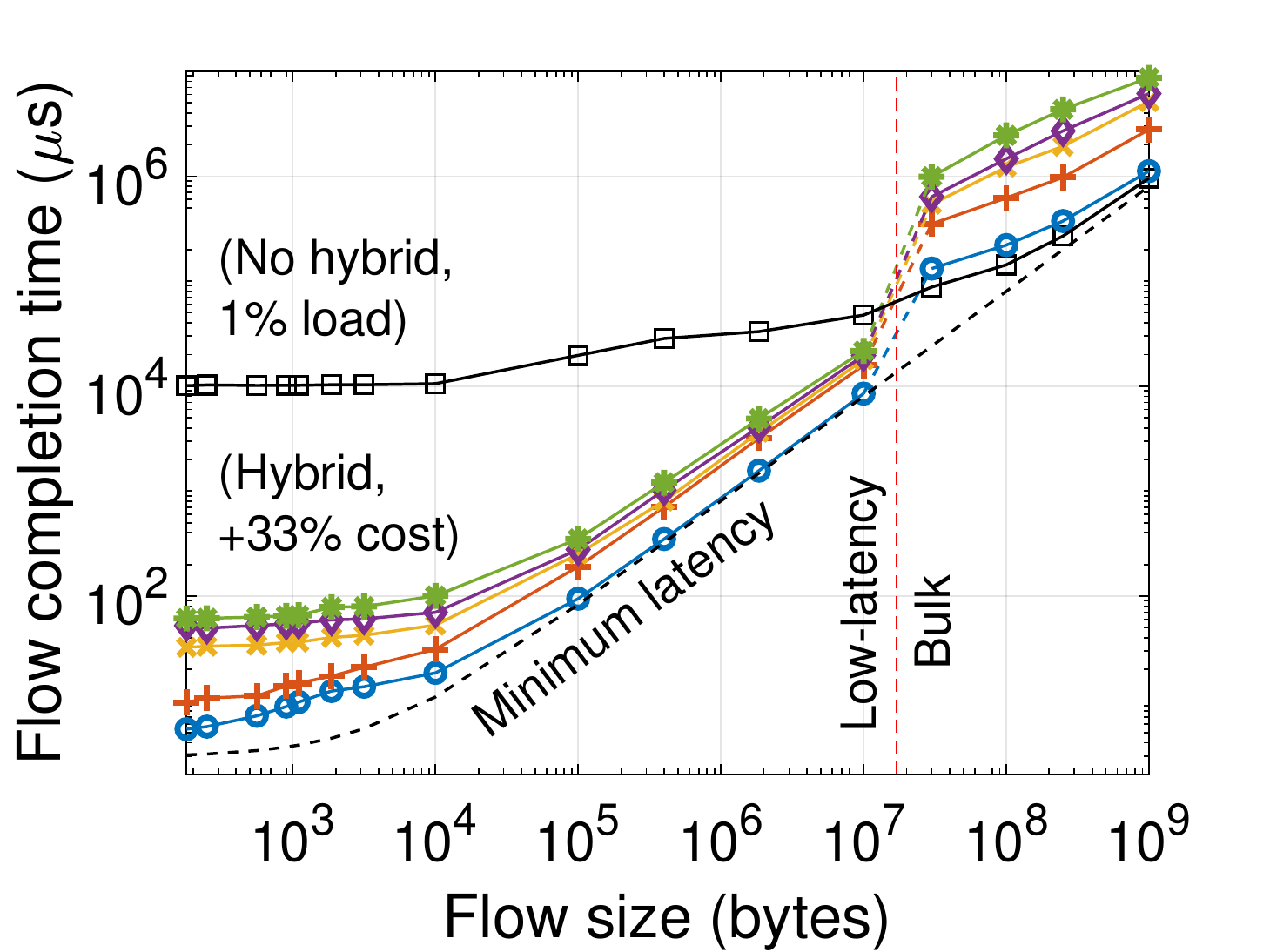}
\caption{\label{fig:fcts_rotornet_1} RotorNet}
\end{subfigure}
\begin{subfigure}[b]{0.24\textwidth}
\includegraphics[width=\linewidth]{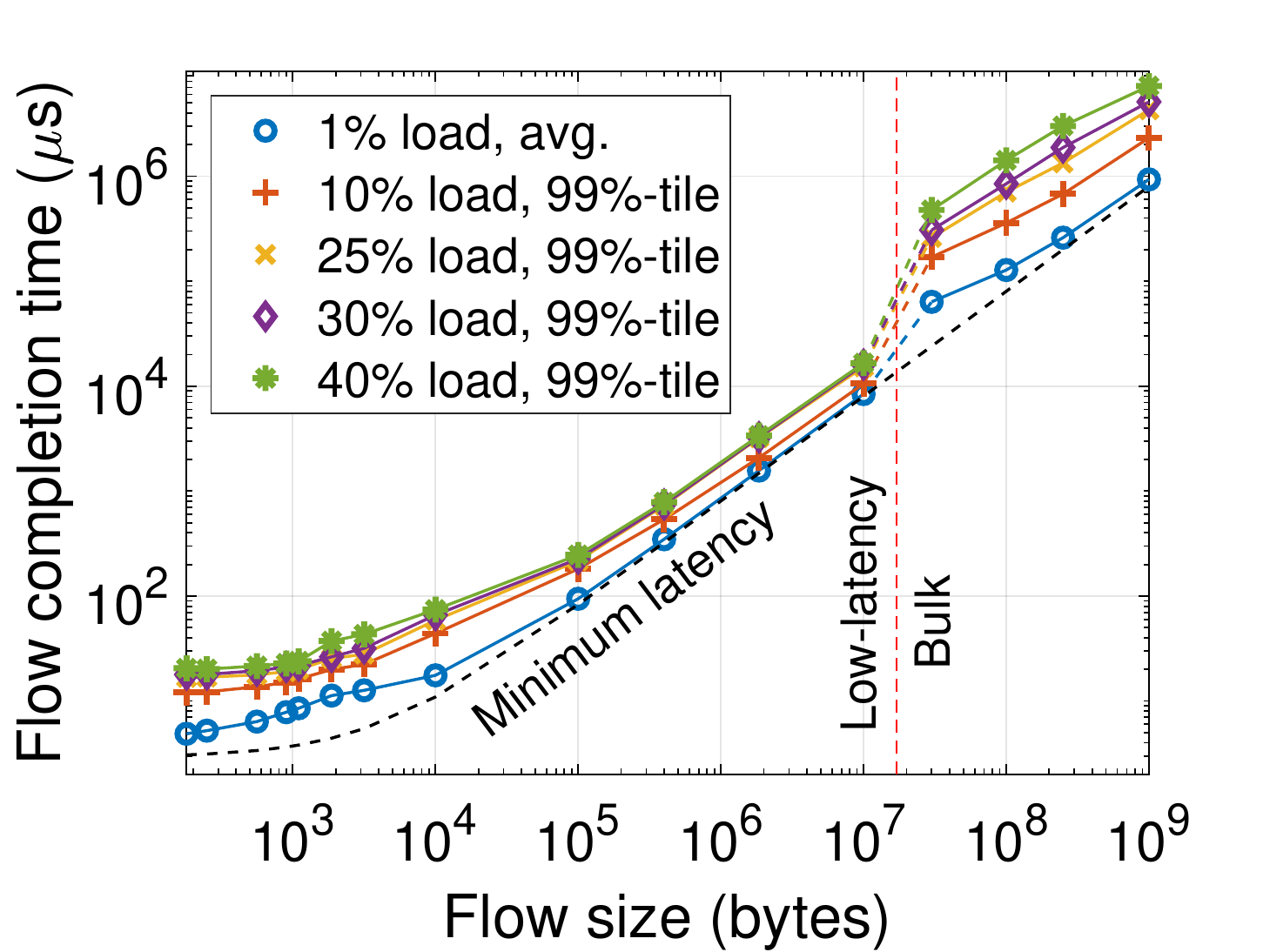}
\caption{\label{fig:fcts_opera_1} Opera}
\end{subfigure}
\caption{\label{fig:datamining} FCTs for the
Datamining workload. All networks are cost comparable except hybrid RotorNet, which is 1.33$\times$ more expensive. In (a) and (b), dashed lines are without priority queuing, and solid lines are with ideal priority queuing.}
\vskip -1em
\end{figure*}

\begin{figure}
\centering
\includegraphics[width=0.9\columnwidth]{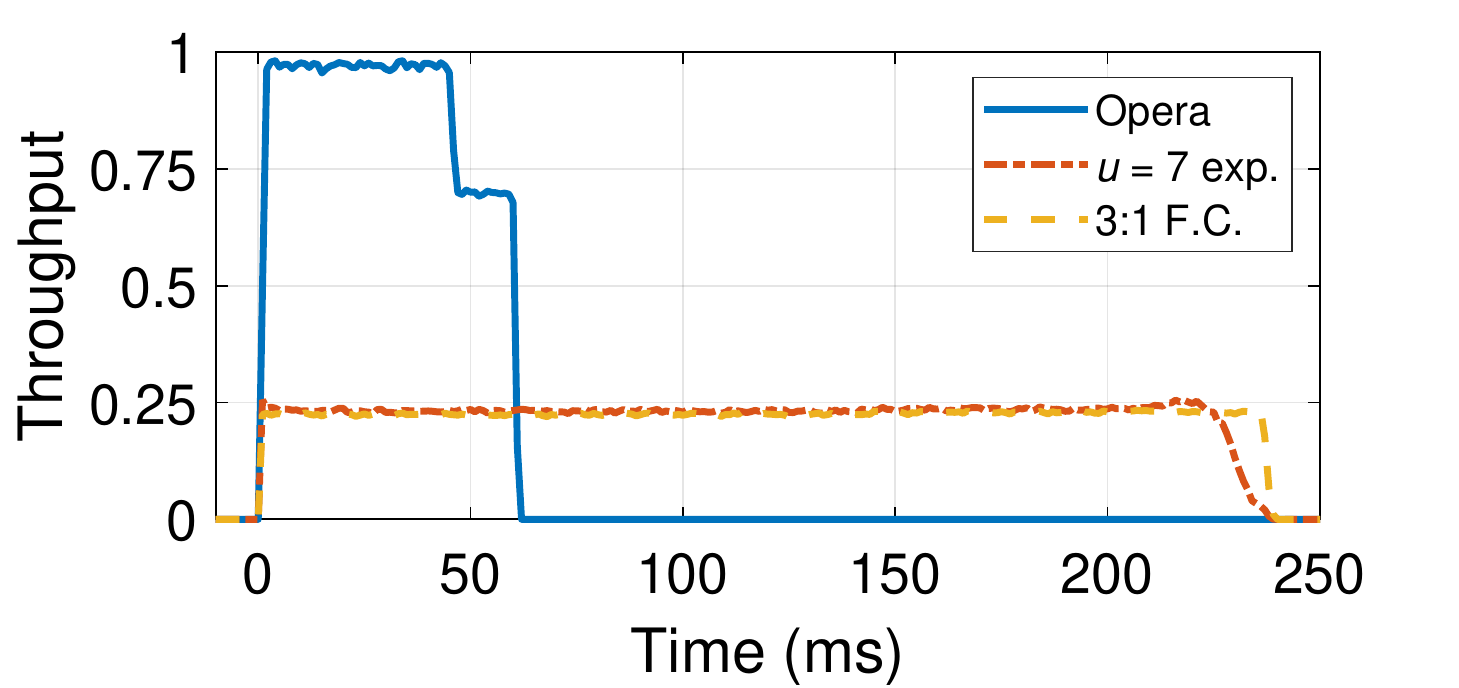}
\caption{\label{fig:tp} Network throughput over time for a 100-KB
  all-to-all Shuffle workload. Opera carries all traffic over direct
  paths, greatly increasing delivered bandwidth. (The small ``step''
  down in Opera's throughput around 50 ms is due to some flows
  finishing in one additional cycle.)}
\vskip -1em
\end{figure}

\begin{figure*}[t]
\centering
\begin{subfigure}[b]{.32\textwidth}
\includegraphics[width=\linewidth]{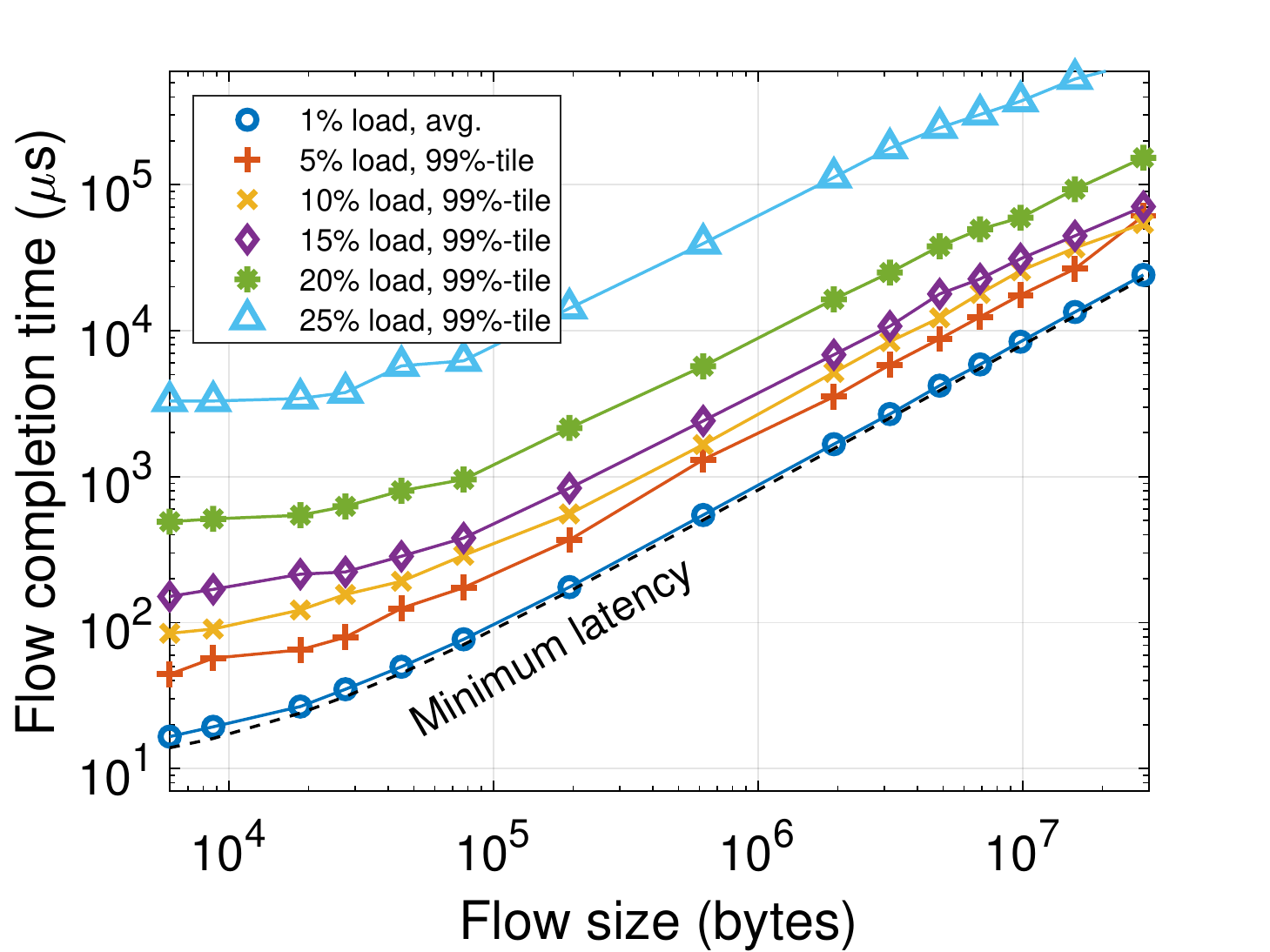}
\caption{\label{fig:fcts_3to1_2} 3:1 folded Clos}
\end{subfigure}
\begin{subfigure}[b]{0.32\textwidth}
\includegraphics[width=\linewidth]{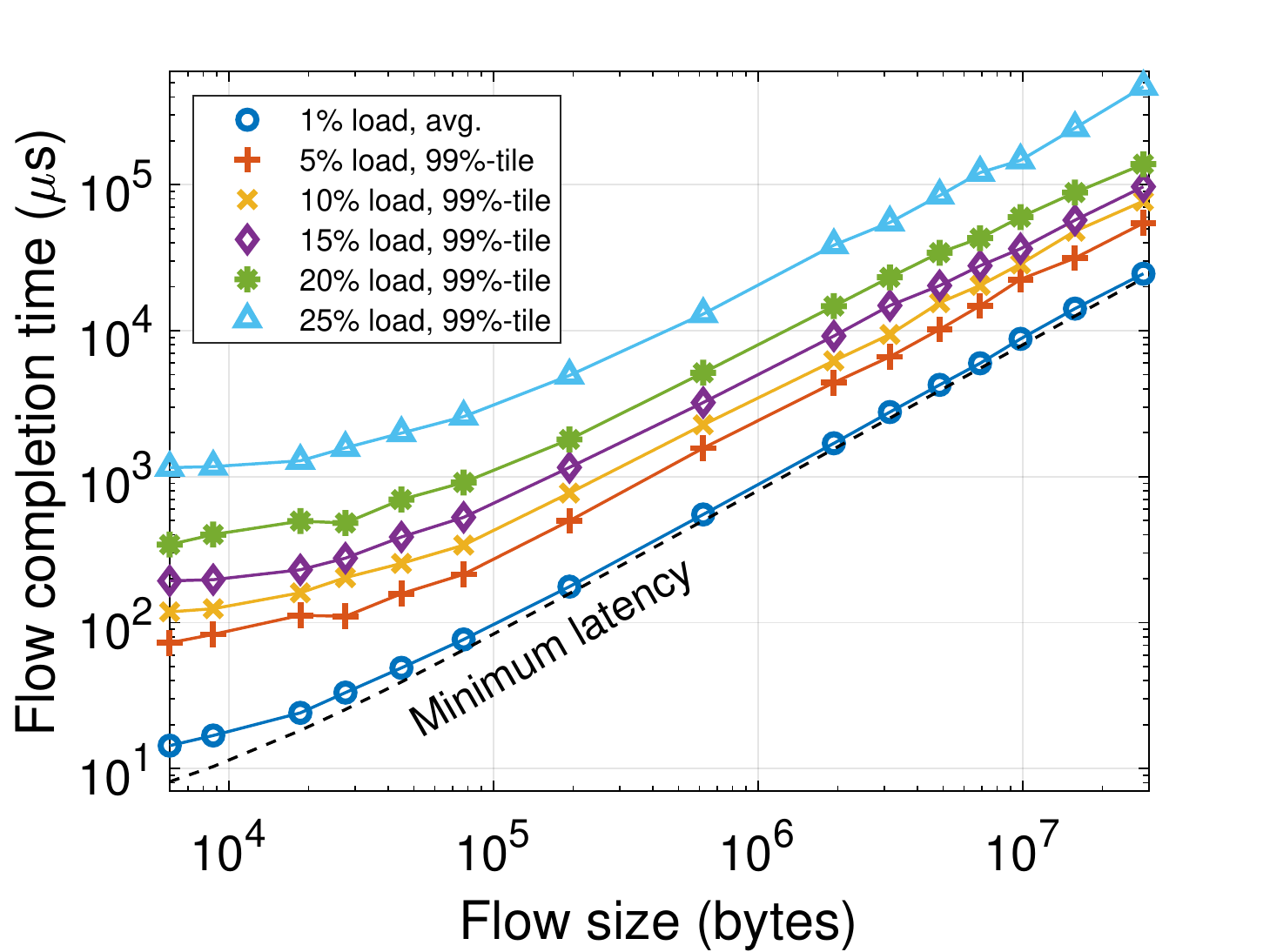}
\caption{\label{fig:fcts_u7_2} $u=7$ expander}
\end{subfigure}
\begin{subfigure}[b]{0.32\textwidth}
\includegraphics[width=\linewidth]{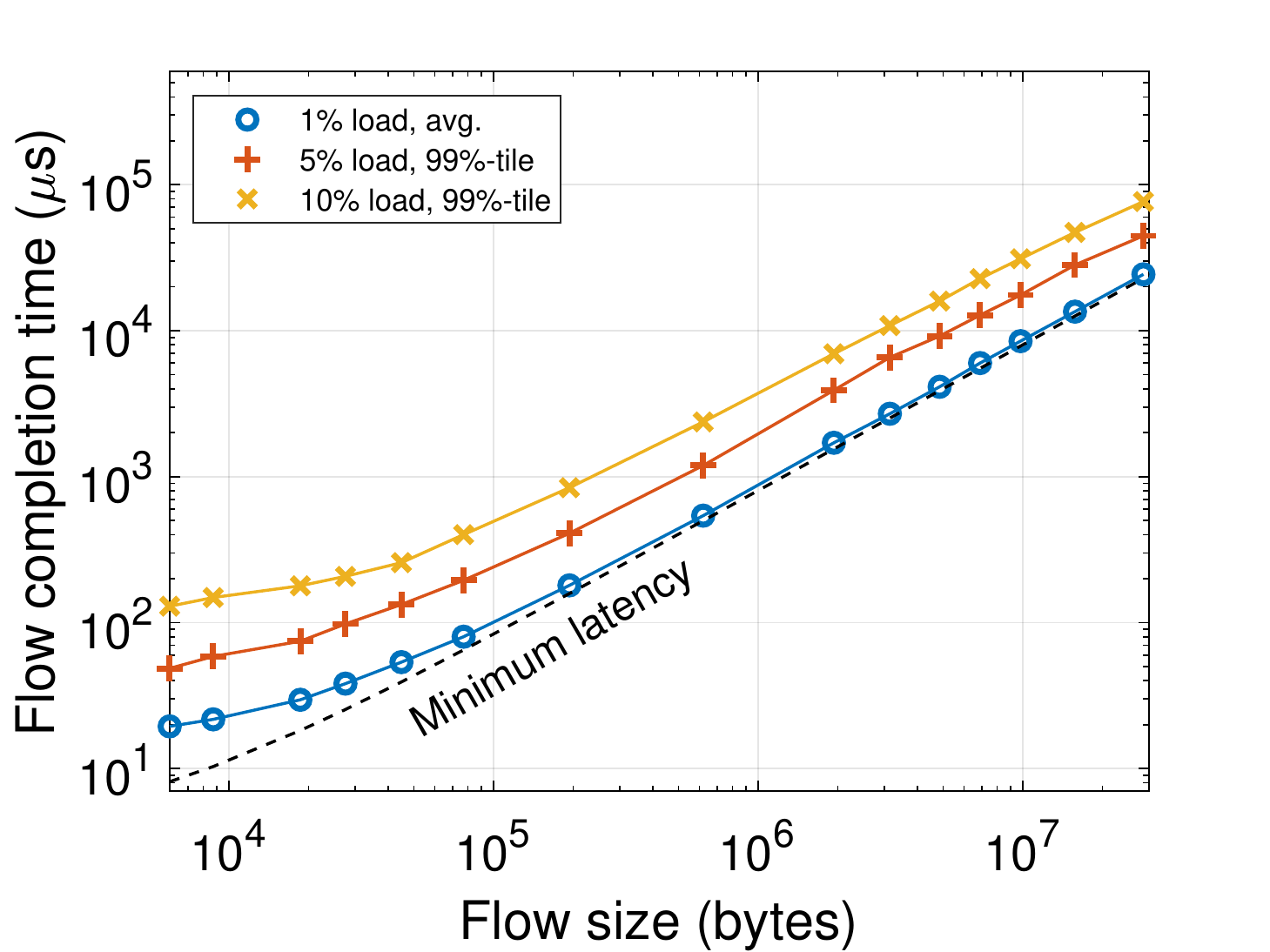}
\caption{\label{fig:fcts_opera_2} Opera}
\end{subfigure}
\caption{\label{fig:websearch} FCTs for the
Websearch workload. Opera carries all traffic over indirect paths.
Opera supports up to 10\% low-latency traffic load with near-equivalent
FCTs to the 3:1 folded Clos and $u=7$ expander.}
\vskip -1em
\end{figure*}

We evaluate Opera in simulation. Initially, we focus on a concrete 648-host
network, comparing to cost-equivalent folded-Clos, static expander, non-hybrid
RotorNet, and (non-cost-equivalent) hybrid RotorNet networks.  We then validate
against a range of network sizes, skewed workloads, and underlying cost
assumptions.
We use the \texttt{htsim} packet
simulator~\cite{htsim_github}, which was previously used to evaluate
the NDP protocol~\cite{ndp}, and extend it to model static expander
networks and dynamic networks. We also modify NDP to handle $<$1500
byte packets, which is necessary for some workloads considered.
Both the folded-Clos and static
expander use NDP as the transport protocol. Opera
and RotorNet use
NDP to
transport low-latency traffic and RotorLB for bulk traffic. Because
Opera explicitly uses priority queuing, we simulate the static networks
with idealized priority queuing where appropriate to maintain a fair
comparison.
Following prior work~\cite{projector,expander2}, we set the link
bandwidth to 10 Gb/s. We use a 1500-byte MTU and set
the propagation delay to 500 ns between ToRs (equivalent to
100 m of fiber).

\subsection{Real-world traffic}

We start by considering Opera's target scenario, a workload with an
inherent mix of bulk and low-latency traffic.  Here we consider the
Datamining workload from Microsoft~\cite{vl2}, and use a Poisson
flow-arrival process to generate flows. We vary the Poisson rate to
adjust the load on the network, defining load relative to the
aggregate bandwidth of all host links (i.e., 100\% load means all
hosts are driving their edge links at full capacity, an inadmissible
load for any over-subscribed network).
As shown in the top portion of Figure~\ref{fig:flowdists}, flows in
this workload range in size from 100 bytes to 1 GB. We use Opera's
default configuration to decide how to route traffic: flows $<$15 MB
are treated as low-latency and are routed over indirect paths,
while flows $\geq$15 MB are treated as bulk and are routed over direct
paths.

Figure~\ref{fig:datamining} shows the performance of Opera as well as
cost-comparable 3:1 folded-Clos and $u=7$ static expander networks for
various offered loads.
We also compared to a hybrid RotorNet which faces one of the six ToR
uplinks to a multi-stage packet switched network to accomodate low-latency
traffic (for 1.33$\times$ the cost), and a cost-equivalent non-hybrid
RotorNet with no packet switching above the ToR.
We report the 99th percentile FCT except in
the case of 1\% load, where the variance in the tail obscures the
trend and so report the average instead.  Note that Opera priority
queues all low-latency flows, while by default the static
networks do not.  For fairness, we also present the
expander and folded Clos with ``ideal'' priority queuing---that is,
removing all flows $\geq$15 MB.
For reference, we also plot the
minimum achievable latency in each network, derived from
the end-to-end delay and link capacity.

The static networks start to saturate past 25\% load: folded Clos have
limited network capacity, and expanders have high bandwidth tax.
Opera, on the other hand, is able to service 40\% load despite having
lower innate capacity than the cost-comparable static expander. Opera
offloads bulk traffic onto bandwidth-efficient paths, and only pays
bandwidth tax on the small fraction (4\%) of low-latency traffic that
transits indirect paths, yielding an effective aggregate bandwidth tax
of 8.4\% for this workload.
Hybrid RotorNet, even with 1/6$^{th}$ of its core capacity packet-switched
(for 33\% higher cost than the other networks), delivers longer flow
completion times than Opera for short flows at loads $>$10\%. A non-hybrid
(i.e. all-optical-core) RototNet would be cost-equivalent to the other networks,
but its latency for short flows would be three orders of magnitude higher
than the other networks, as shown in Figure~\ref{fig:fcts_rotornet_1}.

\subsection{Bulk traffic}

Opera's superiority in the mixed case stems entirely from its ability
to avoid paying bandwidth tax on the bulk traffic.  We highlight
this ability by focusing on a workload in which all flows are routed
over direct paths. We consider an all-to-all shuffle
operation (common to MapReduce style applications), and choose the
flow size to be 100 KB based on the median inter-rack flow size
reported in a Facebook Hadoop cluster~\cite{facebook:sigcomm15}
(c.f. Figure~\ref{fig:flowdists}).  Here we presume the application
tags its flows as bulk, so we do not employ flow-length based
classification; i.e., Opera does not indirect any flows in this
scenario.  We let all flows start simultaneously in Opera, as RotorLB
accommodates such cases gracefully, and stagger flow arrivals over 10
ms for the static networks, which otherwise suffer from severe
startup effects.

Figure~\ref{fig:tp} shows the delivered bandwidth over time for the
different networks. The limited capacity of the 3:1 Clos and high
bandwidth tax rates of the expander significantly extend the FCT of
the shuffle operation, yielding 99th-percentile FCTs of 227 ms and 223
ms, respectively.  Opera's direct paths are bandwidth-tax-free,
allowing higher throughput and reducing the 99th-percentile FCT to 60
ms.

\subsection{Only low-latency flows}
\label{sec:onlyll}

Conversely, workloads in which all flows are routed over indirect
low-latency paths represents the worst case for Opera, i.e., it always
pays a bandwidth tax.  Given our 15 MB threshold for bulk traffic, it
is clear from the bottom portion of Figure~\ref{fig:flowdists} that
the Websearch workload~\cite{dctcp} represents such a case.
A lower threshold would avoid the bandwidth tax, but would
require a shorter cycle time to prevent a significant increase in FCT
for these short ``bulk'' flows.

Figure~\ref{fig:websearch} shows the results for the Websearch workload,
again under a Poisson flow arrival process. All networks provide
equivalent FCTs across all flow sizes for loads at or below 10\%, at
which point Opera is not able to admit additional load.  Both the 3:1
folded Clos and expander saturate (slightly) above 25\% load, but at
that point both deliver FCTs nearly $100\times$ worse than at 1\%
load. While Opera forwards traffic in the same manner as the expander in
this scenario, it has only 60\% of the capacity
and pays an additional 41\%
bandwidth tax due to its longer expected path length.

\subsection{Mixed traffic}

\begin{figure}
  \centering
  \vskip -0.5em
\includegraphics[width=0.9\columnwidth]{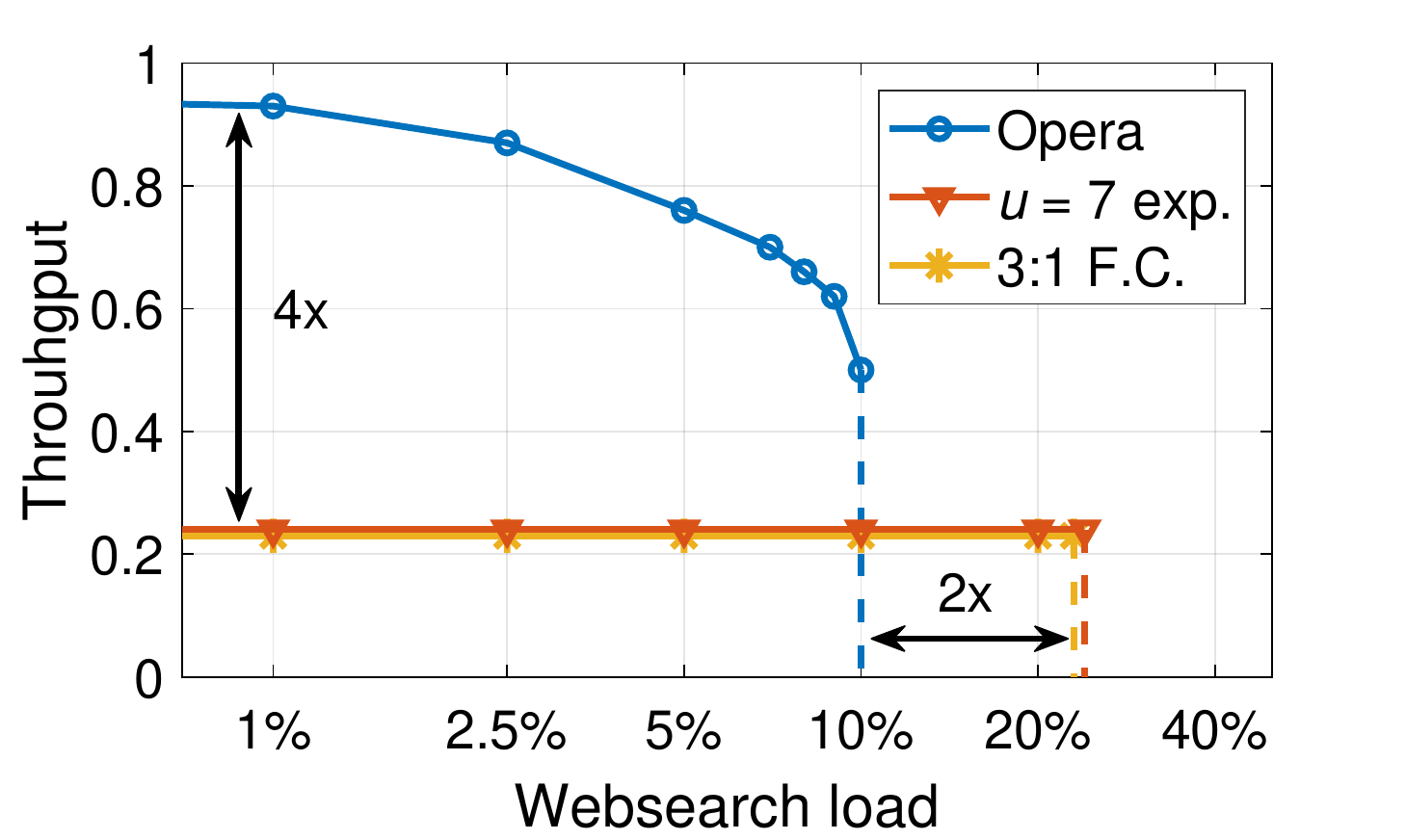}
\caption{\label{fig:tp_v_load} Network throughput vs. Websearch traffic
  load for a combined Websearch/Shuffle workload.}
\vskip -1em
\end{figure}

To drive home Opera's ability to trade off low-latency capacity
against lower effective bandwidth taxes, we explicitly combine the
Websearch (low-latency) and Shuffle (bulk) workloads from above in
varying proportions.
Figure~\ref{fig:tp_v_load} shows the aggregate network throughput as a
function of Websearch (low-latency) traffic load, defined as before as
a fraction of the aggregate host link capacity. We see that for low
Websearch load, Opera delivers up to $4\times$ more throughput than
the static topologies. Even at 10\% Websearch load (near its maximum
admissible load), Opera still delivers almost $2\times$ more
throughput.
In essence, Opera ``gives up'' a factor of 2 in low-latency capacity
(due to its relatively under-provisioned ToRs) to gain a factor of
2--4 in bulk capacity from its vastly lower effective bandwidth tax.

\subsection{Fault tolerance}
\label{sec:eval:ft}

\begin{figure*}
\centering
\includegraphics[width=.32\textwidth]{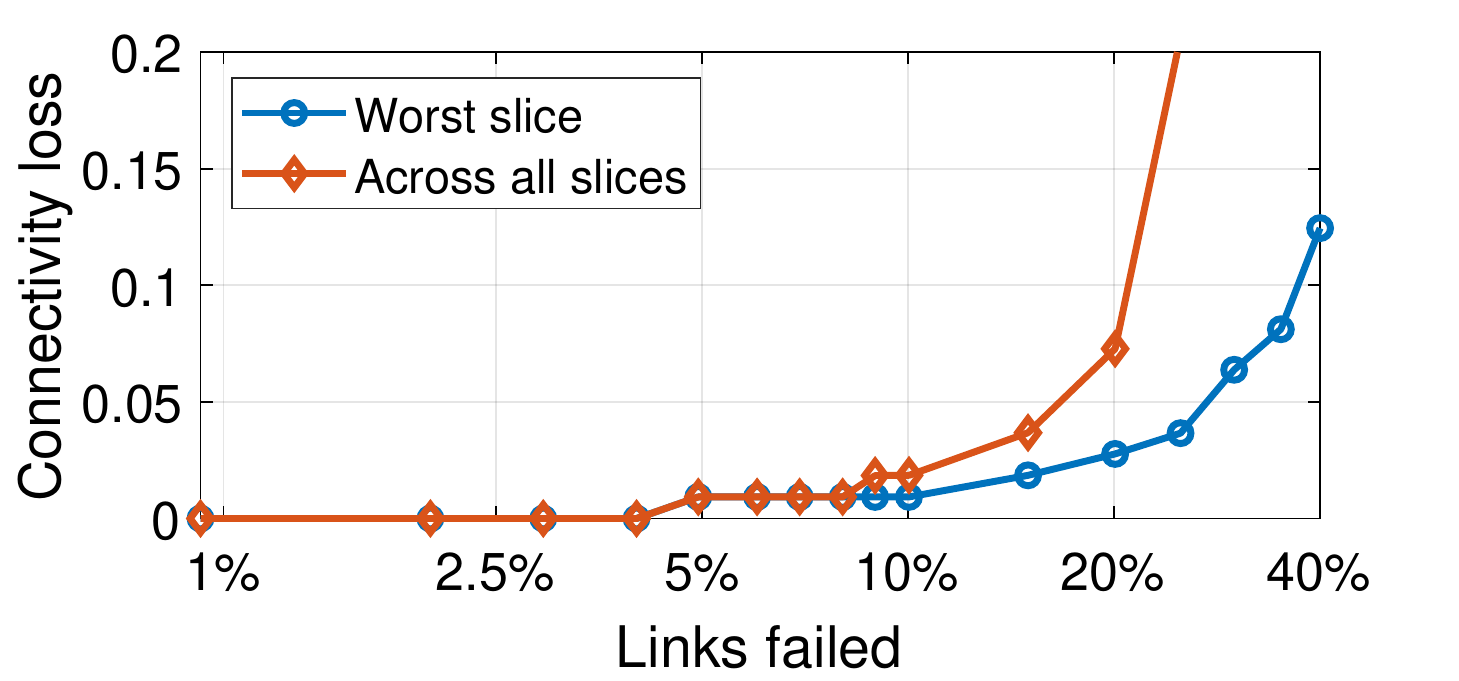}
\includegraphics[width=.32\textwidth]{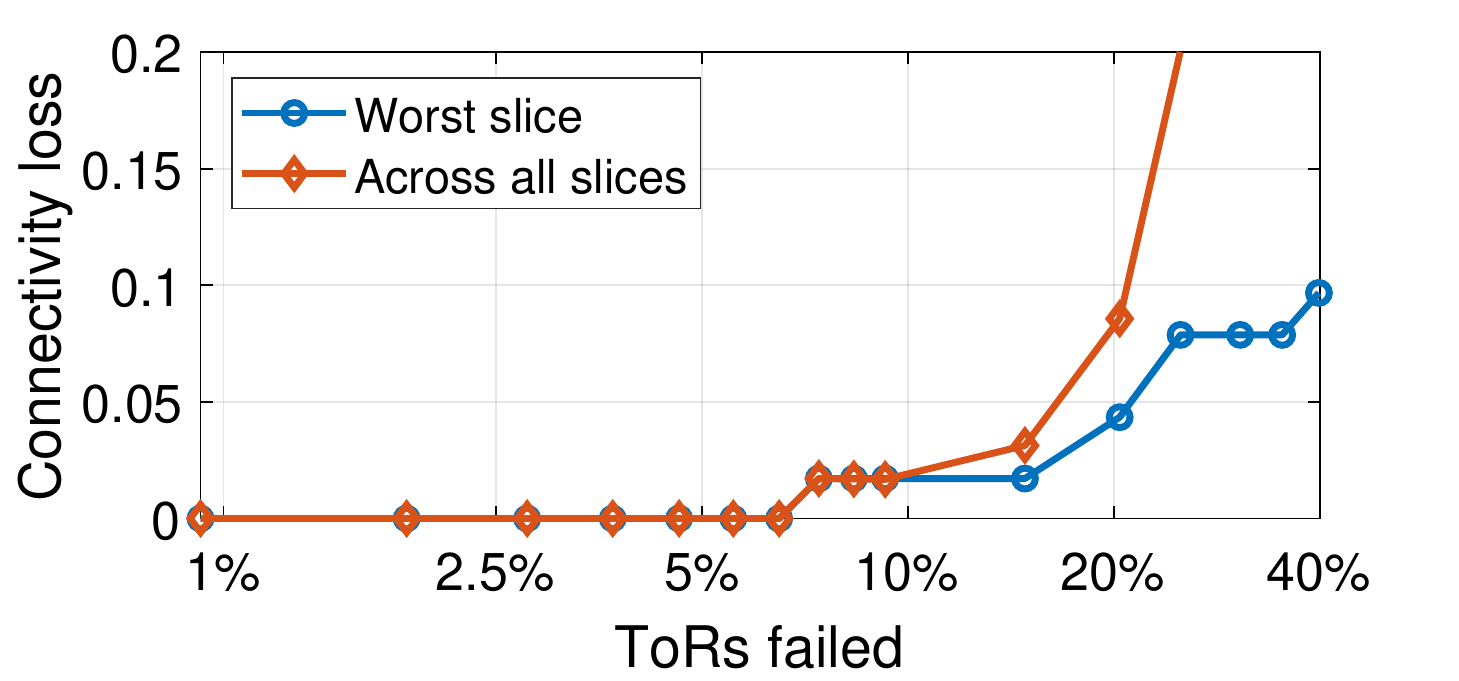}
\includegraphics[width=.32\textwidth]{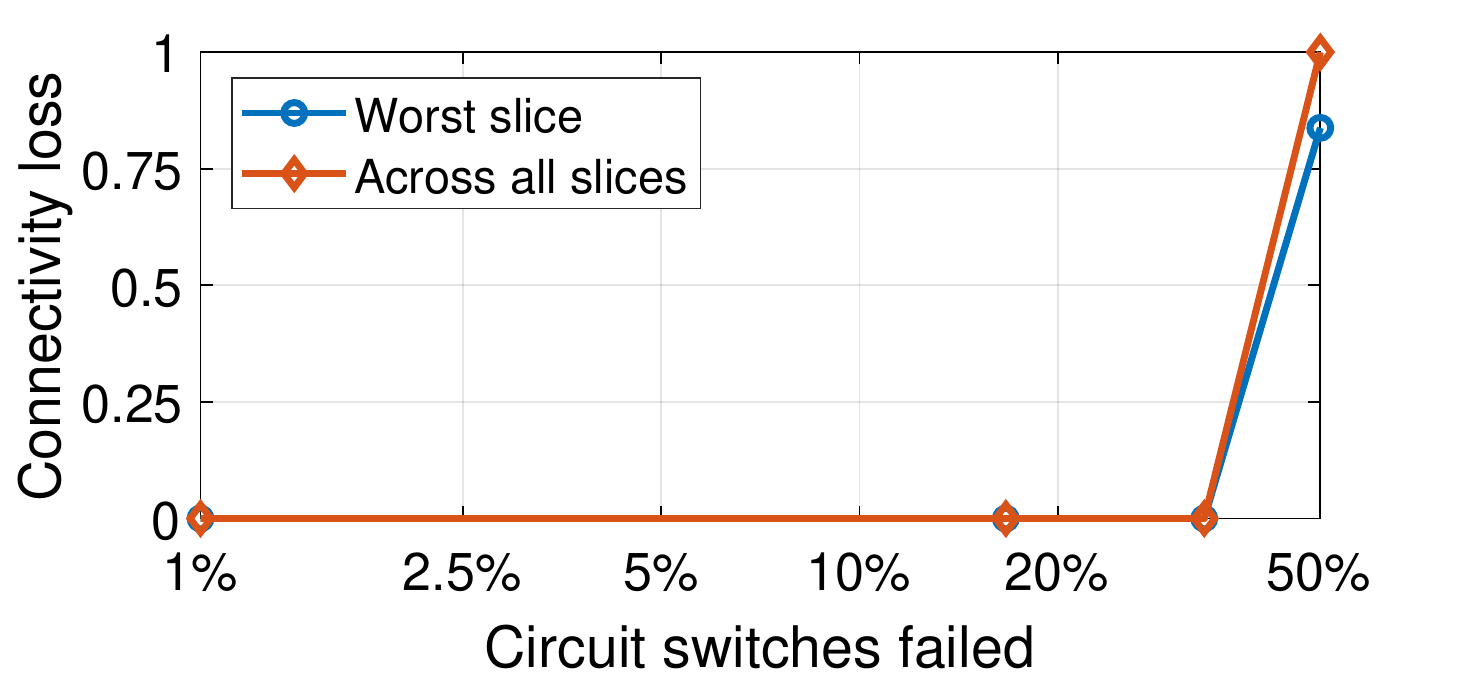}
\caption{\label{fig:fault} Fault tolerance in a 648-host, 108-rack
  Opera network with 6 circuit switches and $k=12$ port
  ToRs. Connectivity loss is the fraction of disconnected ToR
  pairs. In cases involving ToR failures, connectivity loss refers to
  non-failed ToRs.}
\vskip -1em
\end{figure*}

Next, we demonstrate Opera's ability to maintain and re-establish
connectivity in the face of component failures by injecting random
link, ToR, and circuit switch failures into the network. We then
step through the topology slices and record (1) the number of ToR
pairs that were disconnected in the worst-case topology slice and (2)
the number of unique disconnected ToR pairs integrated
across all slices.  Figure~\ref{fig:fault} shows that Opera can
withstand about 4\% of links failing, 7\% of ToRs failing, or 33\% (2
out of 6) of circuit switches failing without suffering any loss in
connectivity. Opera's robustness to failure stems from the good fault
tolerance properties of expander graphs. As discussed in
Appendix~\ref{app:stretch}, Opera has better fault tolerance than a
3:1 folded Clos, and is less tolerant than the $u=7$ expander (which has
higher fanout).
Maintaining connectivity under failure does require some degree of path
stretch in Opera; Appendix~\ref{app:stretch} discusses this in more
detail as well.

\subsection{Network scale and cost sensitivity}
\label{sec:eval:sensitivity}

Finally, we examine Opera's relative performance across a range of
network sizes and cost assumptions. We introduce a parameter
$\alpha$, which is defined following~\cite{expander2} to be the cost
of an Opera ``port'' (consisting of a ToR port, optical transceiver,
fiber, and circuit switch port) divided by the cost of a static network
``port'' (consisting of a ToR port, optical transceiver, and fiber).
A full description of this cost-normalization method is presented in
Appendix~\ref{app:costnormal}. If $\alpha > 1$ (i.e. circuit switch
ports are not free) then a cost-equivalent static network can use the
extra capital to purchase more packet switches and increase its
aggregate capacity.

\begin{figure*}
\centering
\includegraphics[width=.32\textwidth]{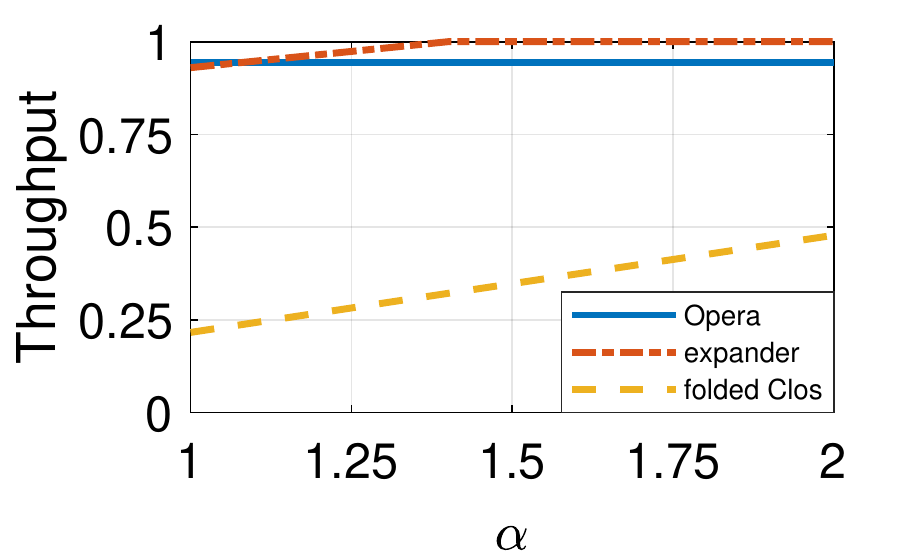}
\includegraphics[width=.32\textwidth]{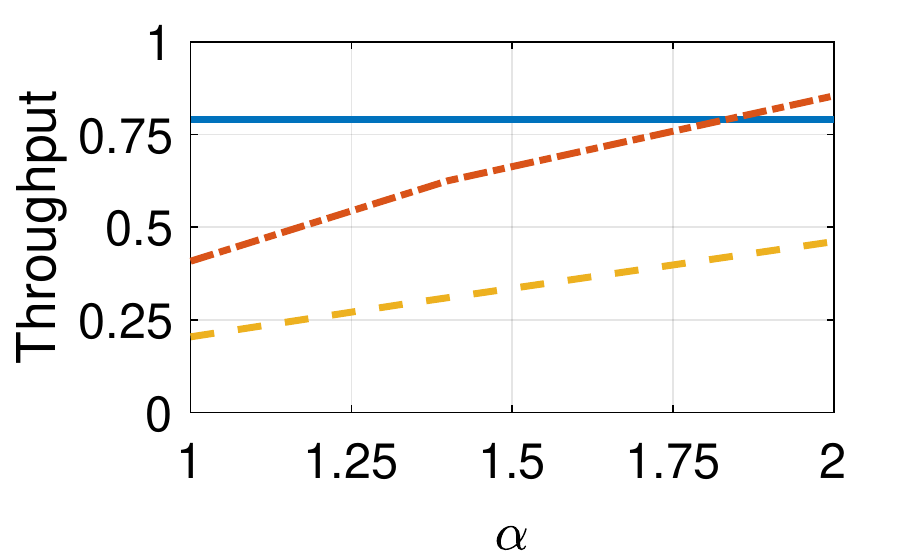}
\includegraphics[width=.32\textwidth]{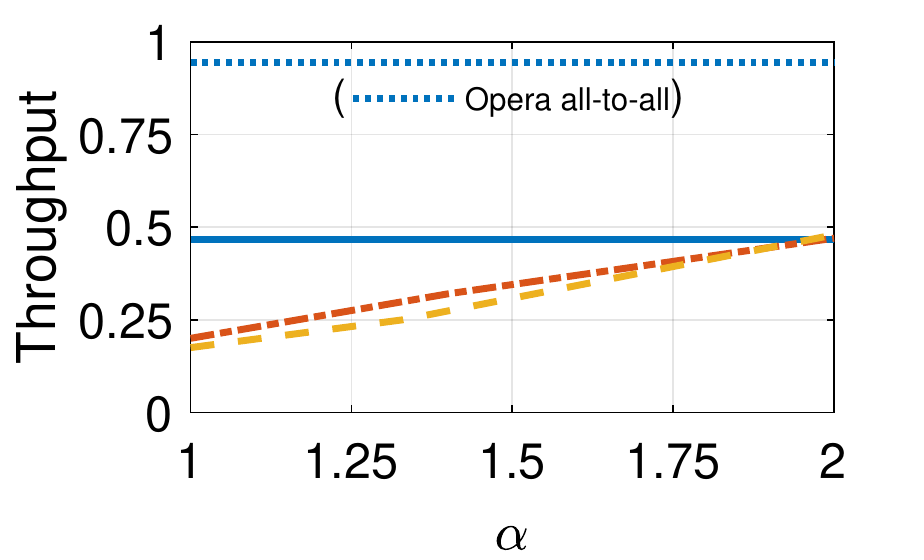}
\caption{\label{fig:throughput:k24} Throughput for (left) hotrack,
  (center) skew[0.2,1], and (right) permutation workloads for $k=24$ ports.}
\vskip -1em
\end{figure*}

We evaluated three workloads using the htsim simulator: (1) hot rack, which
represents a highly skewed workload where one rack communicates with one other
rack; (2) skew[0.2,1], in which 20\% of racks are active (as defined
in~\cite{expander2}); and (3) host permutation, in which each host sends to
one other non-rack-local host.  For each workload we considered a range of
relative Opera port costs (reallocating any resulting cost savings in the
static networks to increase their capacity). We considered both $k=12$ and $k=24$
ToR radices, corresponding to 648-host and 5,184-host networks.
Figure~\ref{fig:throughput:k24} shows the results for $k=24$; the $k=12$ case
has nearly identical performance-cost scaling and is presented in
Appendix~\ref{app:scaling}, along with a path length scaling analysis.

The throughput of the folded Clos topology is independent of traffic pattern,
whereas the throughput of the expander topology decreases as workloads become
less skewed. Opera's throughput initially decreases with a decrease
in skew, then increases as the traffic becomes more uniform.  As long as
$\alpha<1.8$ (Opera’s circuit switch ports costs less than a packet switch port
populated with an optical transceiver), Opera delivers higher throughputs than
either an expander or folded Clos for permutation traffic and moderately skewed
traffic (e.g. 20\% of racks communicating).  In the case of a single hot rack, Opera
offers comperable performance to a static expander.  In the case of shuffle
(all-to-all) traffic, Opera delivers 2$\times$ higher throughput than either the
expander or folded Clos even for $\alpha=2$.

Opera does not offer an advantage for skewed and permutation workloads when the
relative cost of its ports is significantly higher than packet switches
($\alpha>2$), or in deployments where more than 10\% of the link rate is
devoted to urgent, delay-intolerant traffic, as described in
Section~\ref{sec:onlyll}.
 \section{Prototype}
\label{sec:discussion:prototype}
\label{sec:discussion}

Priority queueing plays an important role in Opera's design, ensuring
that low-latency packets do not get buffered behind bulk packets in the
end hosts and switches, and our simulation study reflects this design.
In a real system, low-latency packets that arrive at a switch
might temporarily buffer behind lower-priority bulk packets that are being
transmitted out an egress port.  To better understand the impact of
this effect on the end-to-end latency of Opera,
we built a small-scale hardware
prototype.

The prototype consists of eight ToR switches, each
with four uplinks connected to one of four emulated circuit switches (the same
topology shown in Figure~\ref{fig:topo}).  All eight ToR and four circuit
switches are implemented as virtual switches within a single physical 6.5-Tb/s
Barefoot Tofino switch.
We wrote a P4 program to emulate the circuit switches, which forward
bulk packets arriving at an ingress port
based on a state register, regardless of the destination address of
the packet.  We connect the virtual ToR switches to the four virtual
circuit switches using eight physical 100-Gb/s cables in loopback mode
(logically partitioned into 32 10-Gb/s links).  Each virtual ToR
switch is connected via a cable to one attached end host, which
hosts a Mellanox ConnectX-5 NIC.
There are eight such end hosts (one
per ToR switch) each configured to run at 10 Gb/s.

An attached control server
periodically sends a packet to the Tofino's
ASIC that updates the switch's state register.
After configuring this register, the
controller sends RDMA messages to each of the attached
hosts, signaling that one of the emulated circuit switches has reconfigured.
The end hosts run two processes: an MPI-based shuffle program
patterned on the all-to-all Hadoop workload, and a simple
``ping-pong'' application that sends low-latency RDMA messages to a randomly
selected receiver, which simply returns a response back to the sender.
The relatively low sending rate of the ping-pong application did not require
us to implement NDP for this traffic.

\subsection{End-to-end latency}

\begin{figure}
  \centering
\includegraphics[width=0.9\columnwidth]{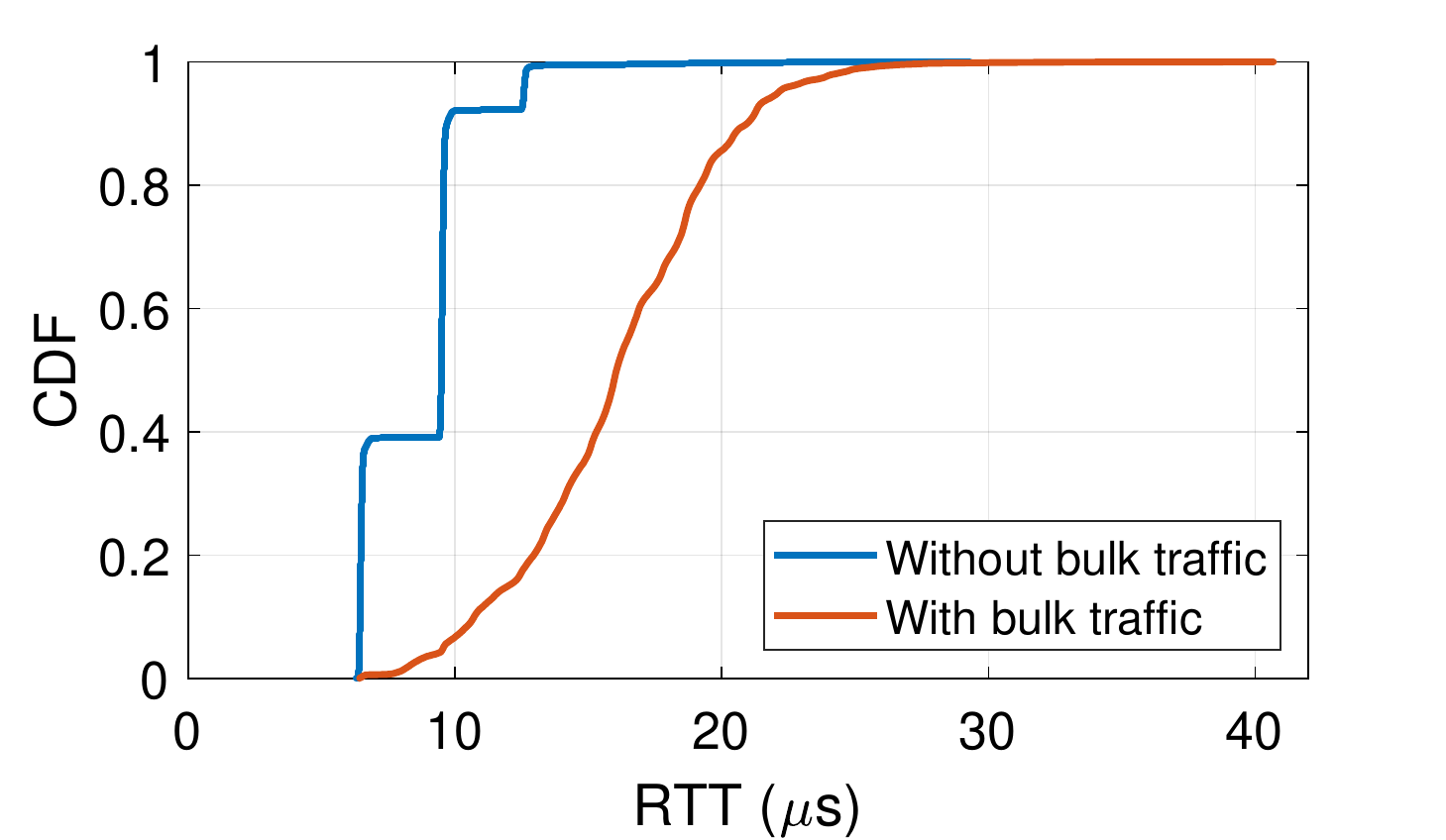}
\caption{\label{fig:proto_result} RTT values
  for low-latency traffic with and without bulk background traffic in the prototype.}
\vskip -1em
\end{figure}

Figure~\ref{fig:proto_result} shows the observed application-level latency
of sending a
ping message from a random source to a random destination (and back).
We plot
this distribution both with and without bulk background traffic.  The
latency observed without bulk traffic is due to a combination of the
path length and the time to forward a packet through Tofino's P4
program, which we observe to be about 3~$\upmu$s per hop, resulting in
latency of up to 9~$\upmu$s depending on path length.  The
observed tail is due to RoCE/MPI variance at the end hosts.
In the presence of bulk traffic, low-latency
packets potentially need
to queue behind bulk packets currently being sent from the egress port.
Because we emulate circuit switches within the Barefoot switch,
each transit of a circuit-switch introduces additional latency that
would not be present in a deployment, adding additional latency.
For our testbed there are as many as eight serialization points from source
to destination, or 16 for each ping-pong exchange.  Each serialization point
can introduce as much as 1.2~$\upmu$s (one MTU at 10 Gb/s), or 19.2~$\upmu$s in total, 
as shown in Figure~\ref{fig:proto_result}.
The distribution is smooth because when low-latency packets buffer
behind bulk packets currently exiting the switch, the amount of remaining
time is effectively a random variable.

\subsection{Routing state scalability}
\label{sec:discussion:feasibility}

\begin{table}[t]
	\centering
	\begin{tabular}{|c|c|c|}
		\hline
		\textbf{\#Racks} & \textbf{\#Entries} & \textbf{\% Utilization} \\ \hline
		108              & 12,096              & 0.7                     \\ \hline
		252              & 65,268              & 3.8                     \\ \hline
		520              & 276,120             & 16.2                    \\ \hline
		768              & 600,576             & 35.3                    \\ \hline
		1008             & 1,032,192            & 60.7                    \\ \hline
		1200             & 1,461,600            & 85.9                    \\ \hline
	\end{tabular}
	\caption{\label{tab:util} Number of entries and resulting resource utilization for Opera rulesets for datacenters of varying sizes.}
\end{table}

Opera
requires more routing state than a static topology.  A
straightforward implementation would require the tables in each switch
to contain $O(N_{rack})^2$ entries as there are $N_{rack}$ topology
slices and $N_{rack}-1$ possible destinations within each slice.
We use Barefoot's
Capilano compiler tool to measure the size of the ruleset for various
datacenter sizes, and compare that size to the capacity of the
Tofino 65x100GE switch.  The ruleset consists of both bulk
and low-latency non-rack-local rules.
The resulting number of rules and the percent utilization of the
switch's memory are shown in Table~\ref{tab:util}.  Because the practical
rulesize limit may be lower than the compiler-predicted size due to
hash collisions within the switch, we loaded the generated rules
into a physical switch to validate that the rules would fit into the
resource constraints.  These results show that today's hardware is
capable of holding the rules needed to implement Opera, while also
leaving spare capacity for additional non-Opera rules.

 \section{Related work}
\label{sec:related}

Opera builds upon previous network designs
focused on cluster and low-latency environments.
In addition to the folded-Clos and expander graph topologies described
thus far, a number of additional static and dynamic network topologies
have been proposed for clusters and datacenters.

\paragraph{Static topologies:} Dragonfly~\cite{dragonfly} and
SlimFly~\cite{slimfly} topologies connect localized pools of high cross-section
bandwidth with a sparse inter-cluster set of links, and have been adopted in
HPC environments.  Diamond~\cite{diamond} and WaveCube~\cite{wavecube}
statically interconnect switches with optical wavelength MUXes, resulting in a
connected topology without reconfiguration.  Quartz~\cite{quartz} interconnects
switches into rings, and relies on multi-hop forwarding
for low-latency traffic.

\paragraph{Dynamic topologies:} Several dynamic network topologies have been
proposed, which we can group into two categories: those that cannot support
low-latency traffic and those that can.  In the former case,
Helios~\cite{helios:sigcomm10}, Mordia~\cite{mordia:sigcomm13}, and
C-Through~\cite{cthrough} aim to reactively establish high-bandwidth
connections in response to observed traffic patterns; they all rely on a
separate packet-switched network to support low-latency traffic.
RotorNet~\cite{rotornet:sigcomm17} relies on deterministic reconfiguration to
deliver constant bandwidth between all endpoints, and then relies on endpoints
injecting traffic using Valiant load balancing to support skewed traffic.
RotorNet requires a separate packet-switched network for low latency traffic.

ProjecToR~\cite{projector}, on the other hand, always maintains a ``base mesh''
of connected links that can handle low-latency traffic while it
opportunistically reconfigures free-space links in response to changes in
traffic patterns.  The authors initially evaluated the use of a random base
network, ruling it out due to poor support of skew.  Instead, they propose a
weighted matching of sources and sinks, though it is not clear what the
expected diameter of that network would be in general.  Like ProjecToR, Opera
maintains an ``always on'' base network which consists of a repeating sequence
of time-varying expander graphs, which has a well-known structure and
performance characteristics.

There are also reconfigurable network proposals that rely on multi-hop
indirection to support low-latency traffic.  In OSA~\cite{OSA}, during
reconfiguration some end-to-end paths may not be available, and so some
circuit-switch ports can be reserved specifically to ensure connectivity for
low-latency traffic.  Megaswitch~\cite{megaswitch} could potentially support
low-latency traffic in a similar manner.

 \section{Conclusions}
\label{sec:conclusions}

Static topologies such as oversubscribed folded-Clos and
expander graphs support low-latency
traffic but have limited overall network bandwidth. Recently proposed dynamic
topologies provide high bandwidth, but cannot support low-latency traffic. 
In this paper, we propose Opera, which is a topology that implements a
series of time-varying expander graphs that support low-latency traffic, and
when integrated over time, provide direct connections between all endpoints to
deliver high throughput to bulk traffic.
Opera can deliver a 4$\times$ increase in throughput for shuffle workloads and
a 60\% increase in supported load for skewed datacenter workloads compared to
cost-equivalent static networks, all without adversely impacting the flow
completion times of short flows.
 
\bibliographystyle{plain}

\begin{thebibliography}{10}

\bibitem{ptp}
{IEEE} standard for a precision clock synchronization protocol for networked
  measurement and control systems.
\newblock {\em {IEEE} Std 1588-2008 (Revision of IEEE Std 1588-2002)}, pages
  1--300, July 2008.

\bibitem{dcswitch:sigcomm08}
Mohammad Al-Fares, Alex Loukissas, and Amin Vahdat.
\newblock A scalable, commodity, data center network architecture.
\newblock In {\em Proceedings of the ACM SIGCOMM Conference}, Seattle, WA,
  August 2008.

\bibitem{hedera:nsdi10}
Mohammad Al-Fares, Sivasankar Radhakrishnan, Barath Raghavan, Nelson Huang, and
  Amin Vahdat.
\newblock {Hedera: Dynamic Flow Scheduling for Data Center Networks}.
\newblock In {\em Proceedings of the 7th ACM/USENIX Symposium on Networked
  Systems Design and Implementation (NSDI)}, San Jose, CA, April 2010.

\bibitem{dctcp}
Mohammad Alizadeh, Albert Greenberg, David~A. Maltz, Jitendra Padhye, Parveen
  Patel, Balaji Prabhakar, Sudipta Sengupta, and Murari Sridharan.
\newblock Data center {TCP} ({DCTCP}).
\newblock In {\em Proceedings of the ACM SIGCOMM Conference}, pages 63--74, New
  Delhi, India, 2010.

\bibitem{hull}
Mohammad Alizadeh, Abdul Kabbani, Tom Edsall, Balaji Prabhakar, Amin Vahdat,
  and Masato Yasuda.
\newblock Less is more: Trading a little bandwidth for ultra-low latency in the
  data center.
\newblock In {\em Proceedings of the 9th USENIX Conference on Networked Systems
  Design and Implementation}, pages 19--19, San Jose, CA, 2012.

\bibitem{alon:expanders}
N~Alon.
\newblock Eigen values and expanders.
\newblock {\em Combinatorica}, 6(2):83--96, January 1986.

\bibitem{pias}
Wei Bai, Li~Chen, Kai Chen, Dongsu Han, Chen Tian, and Hao Wang.
\newblock Information-agnostic flow scheduling for commodity data centers.
\newblock In {\em 12th {USENIX} Symposium on Networked Systems Design and
  Implementation}, pages 455--468, Oakland, CA, 2015.

\bibitem{slimfly}
Maciej Besta and Torsten Hoefler.
\newblock Slim fly: A cost effective low-diameter network topology.
\newblock In {\em Proceedings of the International Conference for High
  Performance Computing, Networking, Storage and Analysis}, pages 348--359, New
  Orleans, Louisana, 2014.

\bibitem{wavecube}
K.~Chen, X.~Wen, X.~Ma, Y.~Chen, Y.~Xia, C.~Hu, and Q.~Dong.
\newblock Wavecube: A scalable, fault-tolerant, high-performance optical data
  center architecture.
\newblock In {\em IEEE Conference on Computer Communications (INFOCOM)}, pages
  1903--1911, April 2015.

\bibitem{OSA}
Kai Chen, Ankit Singlay, Atul Singhz, Kishore Ramachandranz, Lei Xuz, Yueping
  Zhangz, Xitao Wen, and Yan Chen.
\newblock {OSA}: An optical switching architecture for data center networks
  with unprecedented flexibility.
\newblock In {\em Proceedings of the 9th USENIX Conference on Networked Systems
  Design and Implementation}, pages 18--18, San Jose, CA, 2012.

\bibitem{megaswitch}
Li~Chen, Kai Chen, Zhonghua Zhu, Minlan Yu, George Porter, Chunming Qiao, and
  Shan Zhong.
\newblock Enabling wide-spread communications on optical fabric with
  {MegaSwitch}.
\newblock In {\em 14th USENIX Symposium on Networked Systems Design and
  Implementation}, pages 577--593, Boston, MA, 2017.

\bibitem{diamond}
Yong Cui, Shihan Xiao, Xin Wang, Zhenjie Yang, Chao Zhu, Xiangyang Li, Liu
  Yang, and Ning Ge.
\newblock Diamond: Nesting the data center network with wireless rings in {3D}
  space.
\newblock In {\em 13th USENIX Symposium on Networked Systems Design and
  Implementation}, pages 657--669, Santa Clara, CA, 2016.

\bibitem{mapreduce}
Jeffrey Dean and Sanjay Ghemawat.
\newblock {MapReduce}: Simplified data processing on large clusters.
\newblock In {\em Proceedings of the 6th Conference on Symposium on Opearting
  Systems Design \& Implementation}, pages 10--10, San Francisco, CA, 2004.

\bibitem{fctccr}
Nandita Dukkipati and Nick McKeown.
\newblock Why flow-completion time is the right metric for congestion control.
\newblock {\em SIGCOMM Comput. Commun. Rev.}, 36(1):59--62, January 2006.

\bibitem{fb-fabric}
Facebook.
\newblock Facebook's fabric topology.
\newblock
  https://code.facebook.com/posts/360346274145943/introducing-data-center-fabric-the-\-next-generation-facebook-data-center-network/,
  2018.

\bibitem{helios:sigcomm10}
Nathan Farrington, George Porter, Sivasankar Radhakrishnan, Hamid Bazzaz,
  Vikram Subramanya, Yeshaiahu Fainman, George Papen, and Amin Vahdat.
\newblock Helios: A hybrid electrical/optical switch architecture for modular
  data centers.
\newblock In {\em Proceedings of the ACM SIGCOMM Conference}, New Delhi, India,
  August 2010.

\bibitem{Ford:94}
Joseph~E. Ford, Yeshayahu Fainman, and Sing~H. Lee.
\newblock Reconfigurable array interconnection by photorefractive correlation.
\newblock {\em Appl. Opt.}, 33(23):5363--5377, Aug 1994.

\bibitem{hadoop}
The Apache~Software Foundation.
\newblock Apache {Hadoop}.
\newblock https://hadoop.apache.org/, 2018.

\bibitem{phost}
Peter~X. Gao, Akshay Narayan, Gautam Kumar, Rachit Agarwal, Sylvia Ratnasamy,
  and Scott Shenker.
\newblock {pHost}: Distributed near-optimal datacenter transport over commodity
  network fabric.
\newblock In {\em Proceedings of the 11th ACM Conference on Emerging Networking
  Experiments and Technologies}, pages 1:1--1:12, Heidelberg, Germany, 2015.

\bibitem{projector}
Monia Ghobadi, Ratul Mahajan, Amar Phanishayee, Nikhil Devanur, Janardhan
  Kulkarni, Gireeja Ranade, Pierre-Alexandre Blanche, Houman Rastegarfar,
  Madeleine Glick, and Daniel Kilper.
\newblock {ProjecToR}: Agile reconfigurable data center interconnect.
\newblock In {\em Proceedings of the ACM SIGCOMM Conference}, pages 216--229,
  Florianopolis, Brazil, 2016.

\bibitem{vl2}
Albert Greenberg, James~R. Hamilton, Navendu Jain, Srikanth Kandula, Changhoon
  Kim, Parantap Lahiri, David~A. Maltz, Parveen Patel, and Sudipta Sengupta.
\newblock {VL2}: A scalable and flexible data center network.
\newblock In {\em Proceedings of the ACM SIGCOMM Conference on Data
  Communication}, pages 51--62, Barcelona, Spain, 2009.

\bibitem{queuejump}
Matthew~P. Grosvenor, Malte Schwarzkopf, Ionel Gog, Robert N.~M. Watson,
  Andrew~W. Moore, Steven Hand, and Jon Crowcroft.
\newblock Queues don't matter when you can jump them!
\newblock In {\em Proceedings of the 12th USENIX Conference on Networked
  Systems Design and Implementation}, pages 1--14, Oakland, CA, 2015.

\bibitem{firefly}
Navid Hamedazimi, Zafar Qazi, Himanshu Gupta, Vyas Sekar, Samir~R. Das, Jon~P.
  Longtin, Himanshu Shah, and Ashish Tanwer.
\newblock Firefly: A reconfigurable wireless data center fabric using
  free-space optics.
\newblock In {\em Proceedings of the ACM Conference on SIGCOMM}, pages
  319--330, Chicago, Illinois, USA, 2014.

\bibitem{ndp}
Mark Handley, Costin Raiciu, Alexandru Agache, Andrei Voinescu, Andrew~W.
  Moore, Gianni Antichi, and Marcin W\'{o}jcik.
\newblock Re-architecting datacenter networks and stacks for low latency and
  high performance.
\newblock In {\em Proceedings of the Conference of the ACM Special Interest
  Group on Data Communication}, pages 29--42, Los Angeles, CA, USA, 2017.

\bibitem{Hoory06expandergraphs}
Shlomo Hoory, Nathan Linial, and Avi Wigderson.
\newblock Expander graphs and their applications.
\newblock {\em BULL. AMER. MATH. SOC.}, 43(4):439--561, 2006.

\bibitem{htsim_github}
Ht-sim.
\newblock The htsim simulator.
\newblock \url{https://github.com/nets-cs-pub-ro/NDP/wiki/NDP-Simulator}, 2018.

\bibitem{expander1}
Sangeetha~Abdu Jyothi, Ankit Singla, P.~Brighten Godfrey, and Alexandra Kolla.
\newblock Measuring and understanding throughput of network topologies.
\newblock In {\em Proceedings of the International Conference for High
  Performance Computing, Networking, Storage and Analysis}, pages 65:1--65:12,
  Salt Lake City, Utah, 2016.

\bibitem{flyways}
Srikanth Kandula, Jitendra Padhye, and Paramvir Bahl.
\newblock Flyways to de-congest data center networks.
\newblock In {\em Proceedings of the 8th {ACM} Workshop on Hot Topics in
  Networks (HotNets-VIII)}, New York City, NY, October 2009.

\bibitem{expander2}
Simon Kassing, Asaf Valadarsky, Gal Shahaf, Michael Schapira, and Ankit Singla.
\newblock Beyond fat-trees without antennae, mirrors, and disco-balls.
\newblock In {\em Proceedings of the Conference of the ACM Special Interest
  Group on Data Communication}, pages 281--294, Los Angeles, CA, USA, 2017.

\bibitem{dragonfly}
John Kim, Wiliam~J. Dally, Steve Scott, and Dennis Abts.
\newblock Technology-driven, highly-scalable dragonfly topology.
\newblock In {\em Proceedings of the 35th Annual International Symposium on
  Computer Architecture}, pages 77--88, Beijing, China, 2008.

\bibitem{reactor:nsdi14}
He~Liu, Feng Lu, Alex Forencich, Rishi Kapoor, Malveeka Tewari, Geoffrey~M.
  Voelker, George Papen, Alex~C. Snoeren, and George Porter.
\newblock {Circuit Switching Under the Radar with REACToR}.
\newblock In {\em Proceedings of the 11th ACM/USENIX Symposium on Networked
  Systems Design and Implementation (NSDI)}, pages 1--15, Seattle, WA, April
  2014.

\bibitem{quartz}
Yunpeng~James Liu, Peter~Xiang Gao, Bernard Wong, and Srinivasan Keshav.
\newblock Quartz: A new design element for low-latency {DCNs}.
\newblock In {\em Proceedings of the ACM Conference on SIGCOMM}, pages
  283--294, Chicago, Illinois, USA, 2014.

\bibitem{jlt}
W.~M. {Mellette}, G.~M. {Schuster}, G.~{Porter}, G.~{Papen}, and J.~E. {Ford}.
\newblock A scalable, partially configurable optical switch for data center
  networks.
\newblock {\em Journal of Lightwave Technology}, 35(2):136--144, Jan 2017.

\bibitem{rotornet:sigcomm17}
William~M. Mellette, Rob McGuinness, Arjun Roy, Alex Forencich, George Papen,
  Alex~C. Snoeren, and George Porter.
\newblock {RotorNet:} a scalable, low-complexity, optical datacenter network.
\newblock In {\em Proceedings of the ACM SIGCOMM Conference}, Los Angeles,
  California, August 2017.

\bibitem{ptree:hotnets16}
William~M. Mellette, Alex~C. Snoeren, and George Porter.
\newblock {P-FatTree}: A multi-channel datacenter network topology.
\newblock In {\em Proceedings of the 15th ACM Workshop on Hot Topics in
  Networks (HotNets-XV)}, Atlanta, GA, November 2016.

\bibitem{homa}
Behnam Montazeri, Yilong Li, Mohammad Alizadeh, and John Ousterhout.
\newblock Homa: A receiver-driven low-latency transport protocol using network
  priorities.
\newblock In {\em Proceedings of the Conference of the ACM Special Interest
  Group on Data Communication}, pages 221--235, Budapest, Hungary, 2018.

\bibitem{portland}
Radhika Niranjan~Mysore, Andreas Pamboris, Nathan Farrington, Nelson Huang,
  Pardis Miri, Sivasankar Radhakrishnan, Vikram Subramanya, and Amin Vahdat.
\newblock Portland: A scalable fault-tolerant layer 2 data center network
  fabric.
\newblock In {\em Proceedings of the ACM SIGCOMM Conference on Data
  Communication}, pages 39--50, Barcelona, Spain, 2009.

\bibitem{mordia:sigcomm13}
George Porter, Richard Strong, Nathan Farrington, Alex Forencich, Pang-Chen
  Sun, Tajana Rosing, Yeshaiahu Fainman, George Papen, and Amin Vahdat.
\newblock Integrating microsecond circuit switching into the data center.
\newblock In {\em Proceedings of the ACM SIGCOMM Conference}, Hong Kong, China,
  August 2013.

\bibitem{facebook:sigcomm15}
Arjun Roy, Hongyi Zeng, Jasmeet Bagga, George Porter, and Alex~C. Snoeren.
\newblock Inside the social network's (datacenter) network.
\newblock In {\em Proceedings of the ACM SIGCOMM Conference}, London, England,
  August 2015.

\bibitem{shoal}
Vishal Shrivastav, Asaf Valadarsky, Hitesh Ballani, Paolo Costa, Ki~Suh Lee,
  Han Wang, Rachit Agarwal, and Hakim Weatherspoon.
\newblock Shoal: A lossless network for high-density and disaggregated racks.
\newblock Technical report, Cornell, https://hdl.handle.net/1813/49647, 2017.

\bibitem{jupiter}
Arjun Singh, Joon Ong, Amit Agarwal, Glen Anderson, Ashby Armistead, Roy
  Bannon, Seb Boving, Gaurav Desai, Bob Felderman, Paulie Germano, Anand
  Kanagala, Jeff Provost, Jason Simmons, Eiichi Tanda, Jim Wanderer, Urs
  H\"{o}lzle, Stephen Stuart, and Amin Vahdat.
\newblock Jupiter rising: A decade of clos topologies and centralized control
  in google's datacenter network.
\newblock In {\em Proceedings of the ACM Conference on Special Interest Group
  on Data Communication}, pages 183--197, London, United Kingdom, 2015.

\bibitem{jellyfish}
Ankit Singla, Chi-Yao Hong, Lucian Popa, and P.~Brighten Godfrey.
\newblock Jellyfish: Networking data centers randomly.
\newblock In {\em Proceedings of the 9th USENIX Conference on Networked Systems
  Design and Implementation}, pages 17--17, San Jose, CA, 2012.

\bibitem{xpander}
Asaf Valadarsky, Gal Shahaf, Michael Dinitz, and Michael Schapira.
\newblock Xpander: Towards optimal-performance datacenters.
\newblock In {\em Proceedings of the 12th International on Conference on
  Emerging Networking EXperiments and Technologies}, pages 205--219, Irvine,
  California, USA, 2016.

\bibitem{cthrough}
Guohui Wang, David~G. Andersen, Michael Kaminsky, Konstantina Papagiannaki,
  T.S.~Eugene Ng, Michael Kozuch, and Michael Ryan.
\newblock c-through: Part-time optics in data centers.
\newblock In {\em Proceedings of the ACM SIGCOMM Conference}, pages 327--338,
  New Delhi, India, 2010.

\bibitem{3DBeam}
Xia Zhou, Zengbin Zhang, Yibo Zhu, Yubo Li, Saipriya Kumar, Amin Vahdat, Ben~Y.
  Zhao, and Haitao Zheng.
\newblock Mirror mirror on the ceiling: Flexible wireless links for data
  centers.
\newblock In {\em Proceedings of the ACM SIGCOMM Conference on Applications,
  Technologies, Architectures, and Protocols for Computer Communication}, pages
  443--454, Helsinki, Finland, 2012.

\end{thebibliography}

\appendix
\section*{Appendix}
\renewcommand{\thesubsection}{\Alph{subsection}}

\subsection{Cost-normalization approach}
\label{app:costnormal}

In this section, we detail the method we used to analyze a range of
cost-equivalent network topologies at various network scales and technology
cost points.
We begin by defining $\alpha$ as the cost of an Opera ``port''
(consisting of a ToR port, optical transceiver,
fiber, and circuit switch port) divided by the cost of a static network
``port'' (consisting of a ToR port, optical transceiver, and fiber),
following~\cite{expander2}.

We can also interpret $\alpha$ as the cost of the ``core'' ports (i.e. upward-facing
ToR ports and above) per edge port (i.e. server-facing ToR port). Core ports
drive the network cost because they require optical transceivers.
Thus, for a folded Clos we can write $\alpha = 2(T-1)/F$ (where $T$ is the
number of tiers and F is the oversubscription factor). For a static expander,
we can write $\alpha = u/(k-u)$ (where $u$ is the number of ToR uplinks and $k$ is
the ToR radix).

We use a $T=3$ three tier (i.e. three layer) folded Clos as the normalizing
basis and keep the packet switch radix ($k$) and number of hosts ($H$)
constant for each point of network comparison. To determine the number of
hosts as a function of k and $\alpha$, we first solve the for the
oversubscription factor as a function of $\alpha$: $F=2(T-1)/\alpha$
(note $T=3$). Then, we find the number of hosts $H$ in a folded Clos as a
function of $F$, $k$, and $\alpha$: $H=(4F/(F+1))(k/2)^T$ (note $T=3$,
and F is a function of $\alpha$). This allows us to compare networks for
various values of $k$ and $\alpha$, but we also estimate $\alpha$ given
technology assumptions described below.

Opera's cost hinges largely on the circuit switching technology used. While a
wide variety of technologies could be used in principle, using optical rotor
switches~\cite{rotornet:sigcomm17} is likely the most cost-effective because
(1) they provide low optical signal attenuation (about 3 dB)~\cite{jlt}, and
(2) they are compatible with either single mode or multimode signaling by
virtue of their imaging-relay-based design~\cite{jlt}. These factors mean that
Opera can use the same (cost) multimode or simglemode transceivers used in
traditional networks, unlike many other optical network proposals that require
expensive and sophisticated telecom grade gear such as wavelength tunable
transceivers or optical amplifiers. Based on the cost estimates of commodity
components taken from~\cite{expander2} and rotor switch
components (summarized in Table~\ref{tbl:partcosts}), we approximate that an
Opera port costs about 1.3$\times$ more than a static network port
(i.e. $\alpha$=1.3).

\begin{table}[]
\centering
\begin{tabular}{l l l }
\cline{1-3}
Component & Static & Opera \\ \hline
\multicolumn{1}{l}{SR transceiver}          & \$80                & \$80
\\ \multicolumn{1}{l}{Optical fiber (\$0.3/m)} & \$45                & \$45
\\ \multicolumn{1}{l}{ToR port}                & \$90                & \$90
\\ \multicolumn{1}{l}{Optical fiber array}     & -                   & \$30 $\dagger$
\\ \multicolumn{1}{l}{Optical lenses}          & -                   & \$15 $\dagger$
\\ \multicolumn{1}{l}{Beam-steering element}   & -                   & \$5 $\dagger$
\\ \multicolumn{1}{l}{Optical mapping}         & -                   & \$10 $\dagger$
\\ \hline
\multicolumn{1}{l}{Total}                   & \$215               & \$275
\\
\multicolumn{1}{l}{$\alpha$ ratio}          & 1              & 1.3
\\ \hline
\end{tabular}
\caption{\label{tbl:partcosts}
Cost per ``port'' for a static network vs. Opera. A ``port'' in a static network
consists of a packet switch port, optical transceiver, and fiber. A ``port'' in
Opera consists of a packet switched (ToR) port, optical transceiver, and fiber,
as well as the components needed to build a rotor switch. The cost of rotor
switch components is amortized across the number of ports on a given rotor switch,
which can be 100s or 1,000s; we present values in the table assuming 512 port
rotor switches. ($\dagger$ per duplex fiber port)}
\end{table}

\subsection{Reducing cycle time at scale}
\label{app:cycle}

\begin{figure}
\centering
\includegraphics[width=0.9\columnwidth]{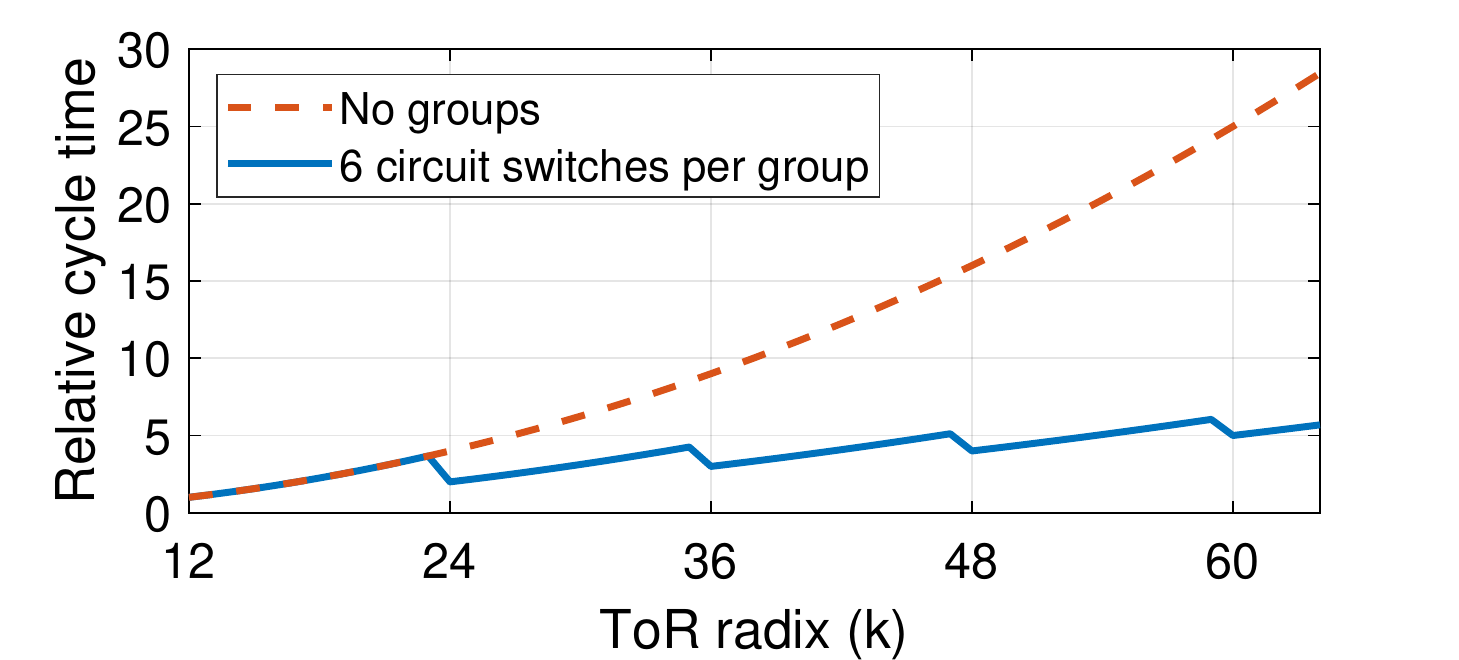}
\caption{\label{fig:cycle_scaling} Relative cycle time is improved at larger
scale by grouping circuit switches and allowing one switch in each group to
reconfigure simultaneously.}
\end{figure}

Larger Opera networks are enabled by higher radix ToR switches, which
commensurately increase the number of circuit switches. To prevent the
cycle time from scaling quadratically with the ToR radix, we allow
multiple circuit switches to reconfigure simultaneously (ensuring that
the remaining switches deliver a fully-connected network at all times).
As an example, doubling the ToR radix doubles the number of circuit
switches, but presents the opportunity to cut the cycle time in half by
reconfiguring two circuit switches at a time. This approach offers linear
scaling in the cycle time with the ToR radix, as shown in
Figure~\ref{fig:cycle_scaling}. Assuming we divide circuit switches into
groups of 6, parallelizing the cycle of each group, the cycle time
increases by a factor of 6 from a $k=12$ (648-host network) to a $k=64$ (98,304-host network), corresponding to a flow length cutoff for ``bulk'' flows of
90 MB in the latter case.

\subsection{Additional scaling analysis}
\label{app:scaling}

\begin{figure*}
\centering
\includegraphics[width=.32\textwidth]{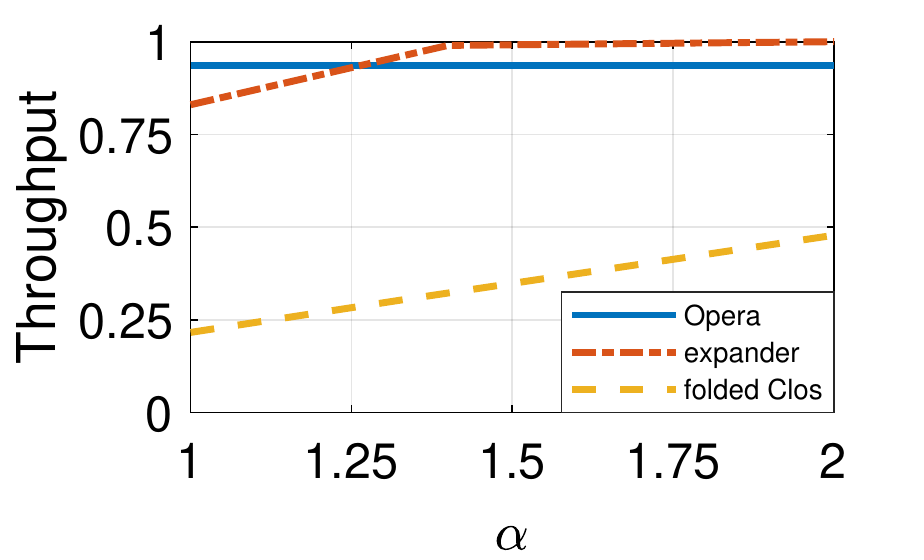}
\includegraphics[width=.32\textwidth]{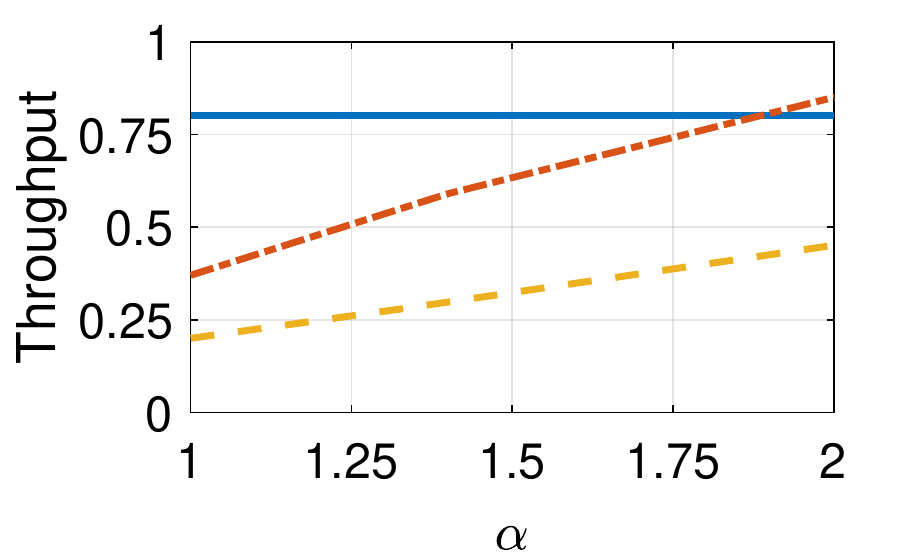}
\includegraphics[width=.32\textwidth]{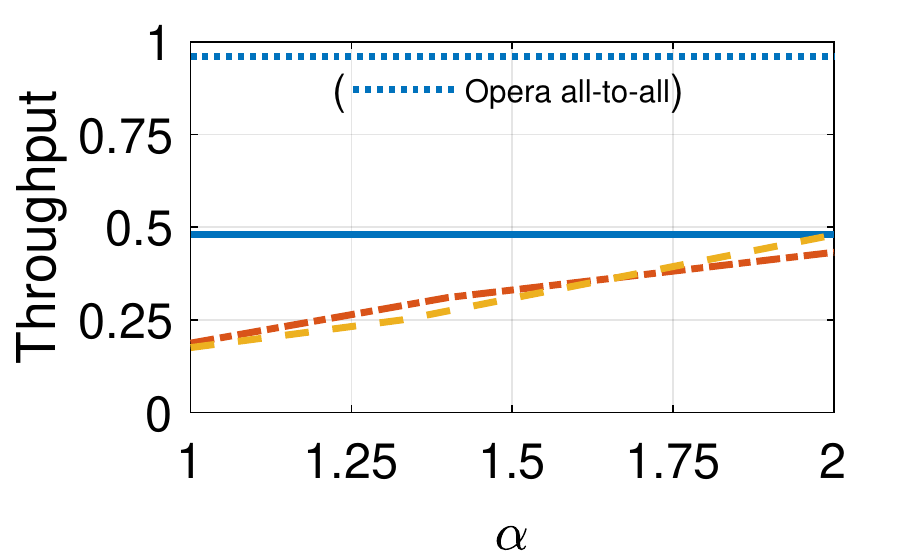}
\caption{\label{fig:throughput:k12} Throughput for (left) hotrack,
(center) skew[0.2,1], and (right) permutation workloads for $k=24$ ports.}
\end{figure*}

\begin{figure}
  \centering
\includegraphics[width=0.9\columnwidth]{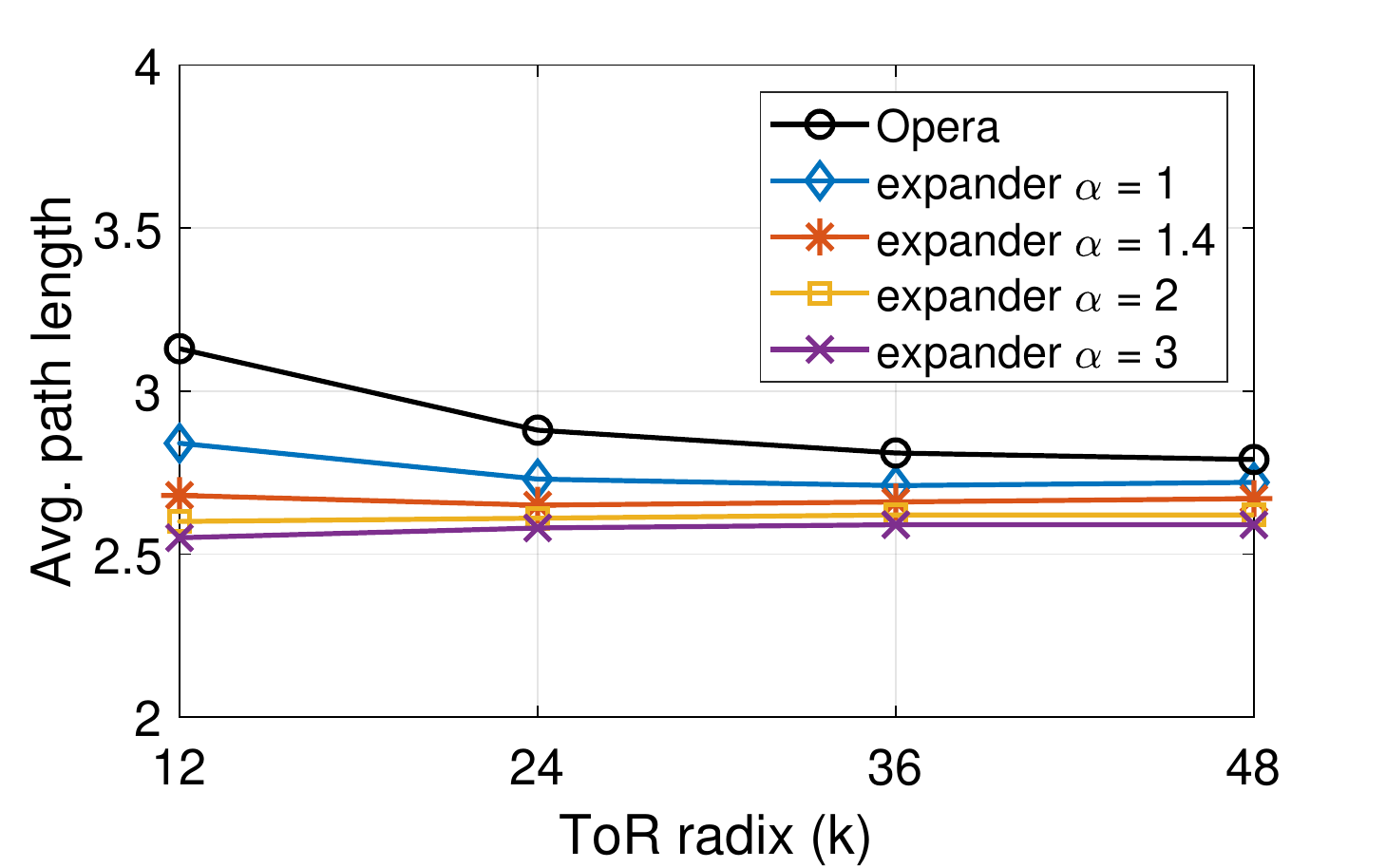}
\caption{\label{fig:hop_scaling} Path lengths for different
network sizes (from $k=12$ with $\approx$ 650 hosts to $k=48$ with $\approx$
98,000 hosts) and relative cost assumptions ($\alpha$).}
\end{figure}

Figure~\ref{fig:throughput:k12} shows the performance-cost scaling trends for
various traffic patterns for networks with $k=12$ port ToRs. Comparing with 
Figure~\ref{fig:throughput:k24}, we observed nearly
indentical performance between networks with $k=12$ and $k=24$, indicating the
(cost-normalized) network performance is nearly independent of scale for all
networks considered (folded Clos, static expanders, and Opera).

To analyze this result at a more fundamental level, we evaluated the average and
worst-case path lengths
for ToR radices between $k=12$ and $k=48$ for both Opera and static expanders
at various cost points ($\alpha$). Figure~\ref{fig:hop_scaling} shows that the average
path lengths converge for large network sizes (the worst-case path length for
all networks including Opera was 4 ToR-to-ToR hops for $k=24$ and above). Given that
the network performance properties of static expanders are correlated with their
path length properties~\cite{alon:expanders}, Figure~\ref{fig:hop_scaling} supports
our observation that the cost-performance properties of the networks do not
change substantially with network size.

\subsection{Spectral efficiency and path lengths}
\label{app:graph}

\begin{figure}
\centering
\includegraphics[width=0.75\columnwidth]{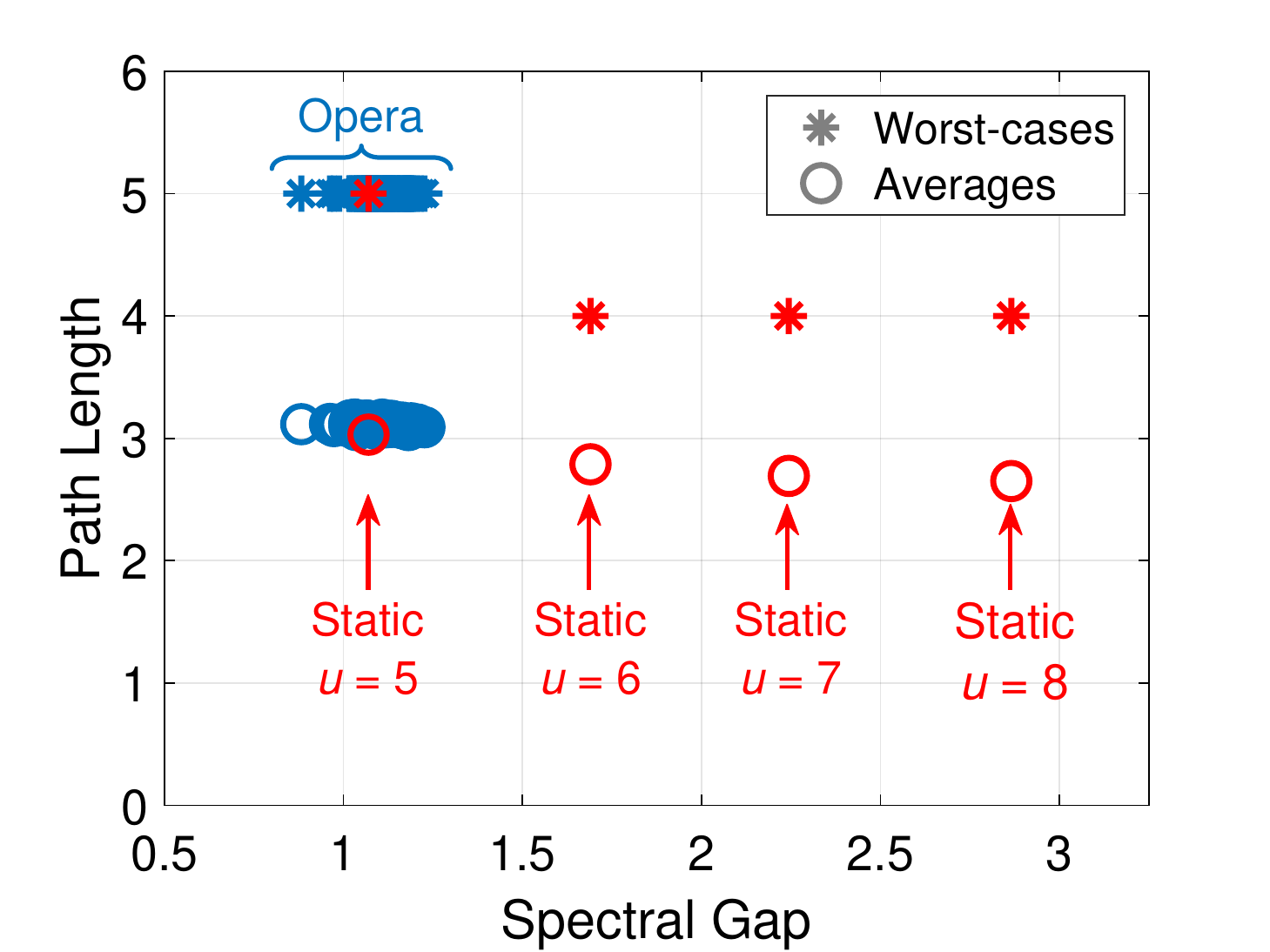}
\caption{\label{fig:hops1} Average and worst-case path lengths and
  spectral gap for Opera and static expander networks. All networks
  use $k=12$-port ToR switches and have between 644 and 650 hosts. Each
  data point for Opera corresponds to one of its 108 topology slices.}
\end{figure}

The \textit{spectral gap} of a network is a graph-theoretic metric indicating
how close a graph is to an optimal Ramanujan expander~\cite{Hoory06expandergraphs}.
Larger spectral gaps imply better expansion. We evaluated the spectral gap
for each the $108$ topology slices in the example 648-host 108-rack Opera
network analyzed in the text, and compared it to the spectral gaps of a number
or randomly-generated static expanders with varying $d$:$u$ ratios. All networks
used $k=12$ radix ToRs and were constrained to have a nearly-equal number of
hosts. The results are shown in Figure~\ref{fig:hops1}.
Note that expanders with larger $u$ require more ToR switches (i.e., cost more)
to support the same number of hosts.

Interestingly, when the number of hosts is held constant, we observe
that the average and worst-case path length is not a strong function
of the spectral gap. Further, we see that Opera comes very close to
the best average path length achievable with a static expander,
indicating that it makes good use of the ToR uplinks in each topology
slice. Opera achieves this good performance despite the fact that we
have imposed additional constraints to support bulk traffic with low
bandwidth tax:
unlike a static expander, Opera must provide a set of $N_{racks}=108$
expanders across time, and those expanders are constructed from an
underlying set of disjoint matchings.

\subsection{Additional failure analysis}
\label{app:stretch}

\begin{figure*}
\centering
\includegraphics[width=.32\textwidth]{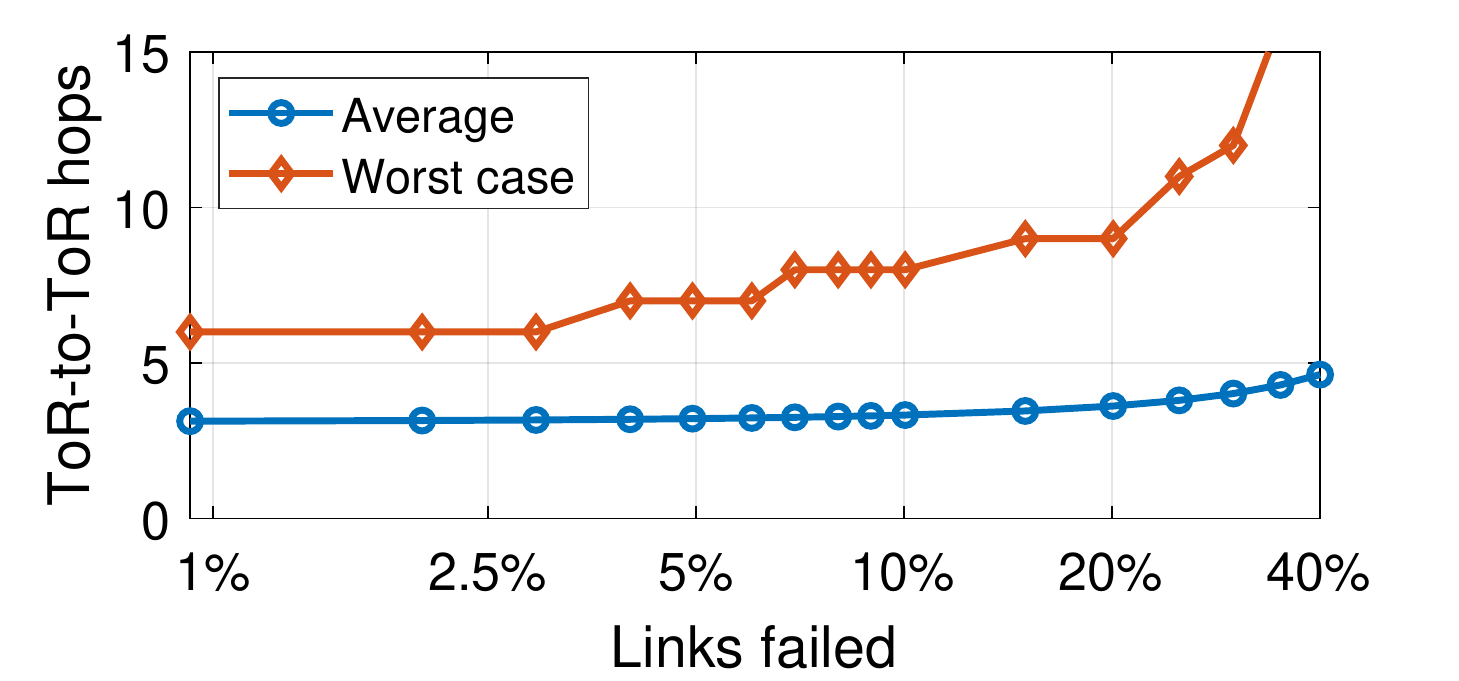}
\includegraphics[width=.32\textwidth]{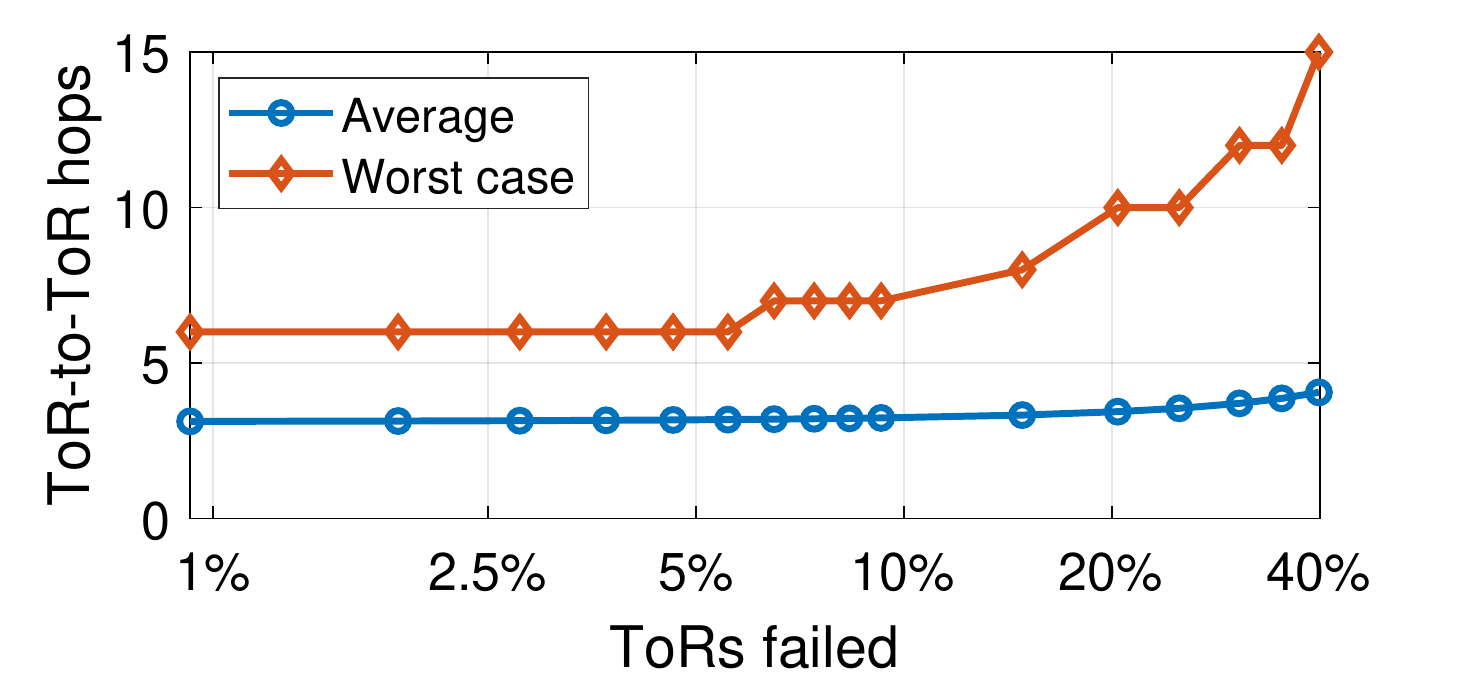}
\includegraphics[width=.32\textwidth]{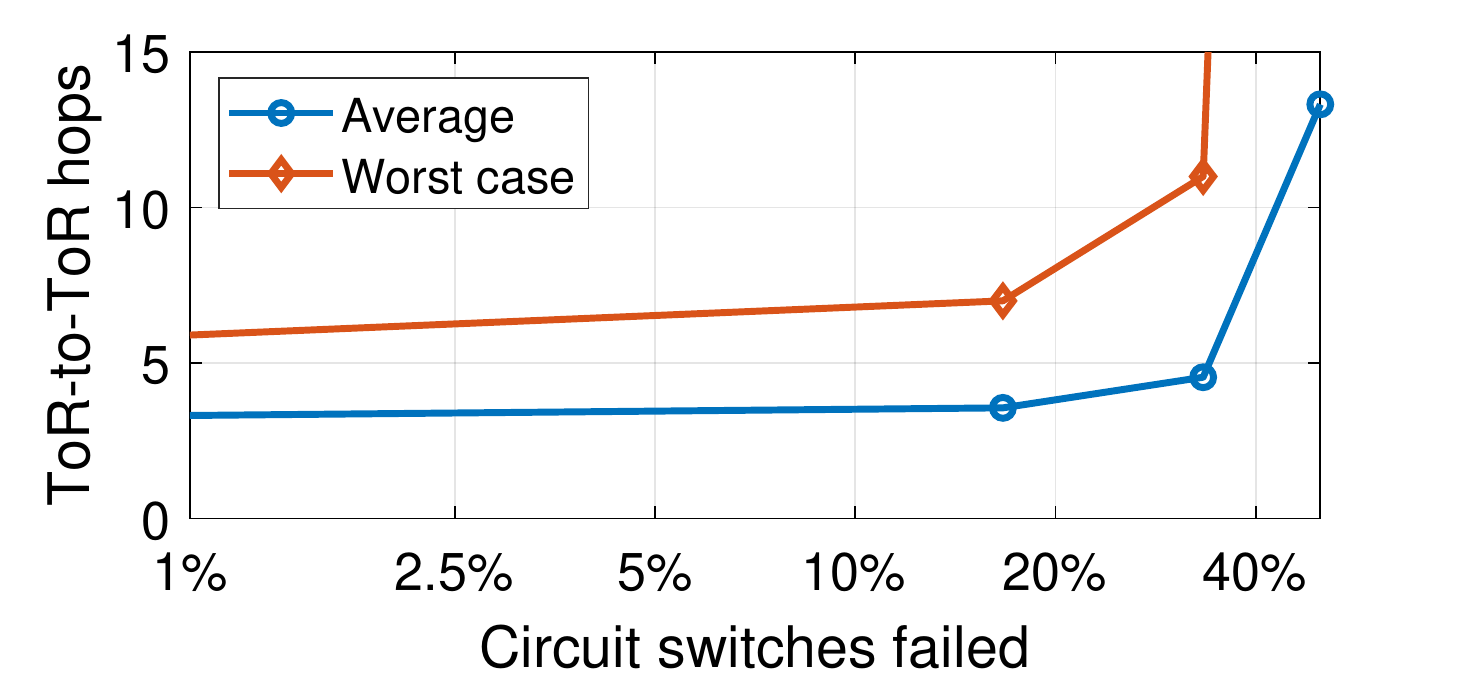}
\caption{\label{fig:stretch} Average and worst-case path length of a 108-rack Opera
network with 6 circuit switches and $k=12$ port ToRs, for various failure conditions.
Path length is reported for all finite-length paths. Figure~\ref{fig:fault} indicates
how many ToR-pairs are disconnected (i.e. have infinite path length).}
\end{figure*}

\begin{figure*}
\centering
\includegraphics[width=.33\textwidth]{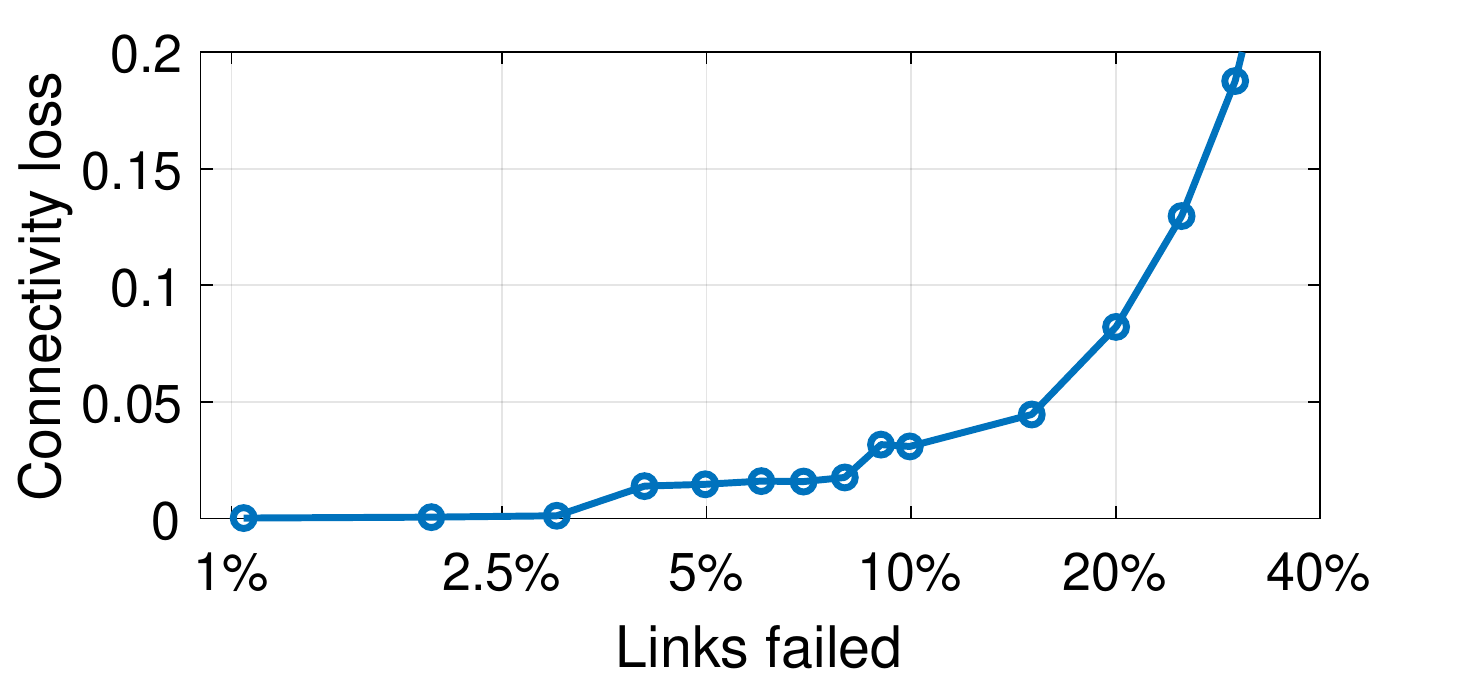}
\includegraphics[width=.33\textwidth]{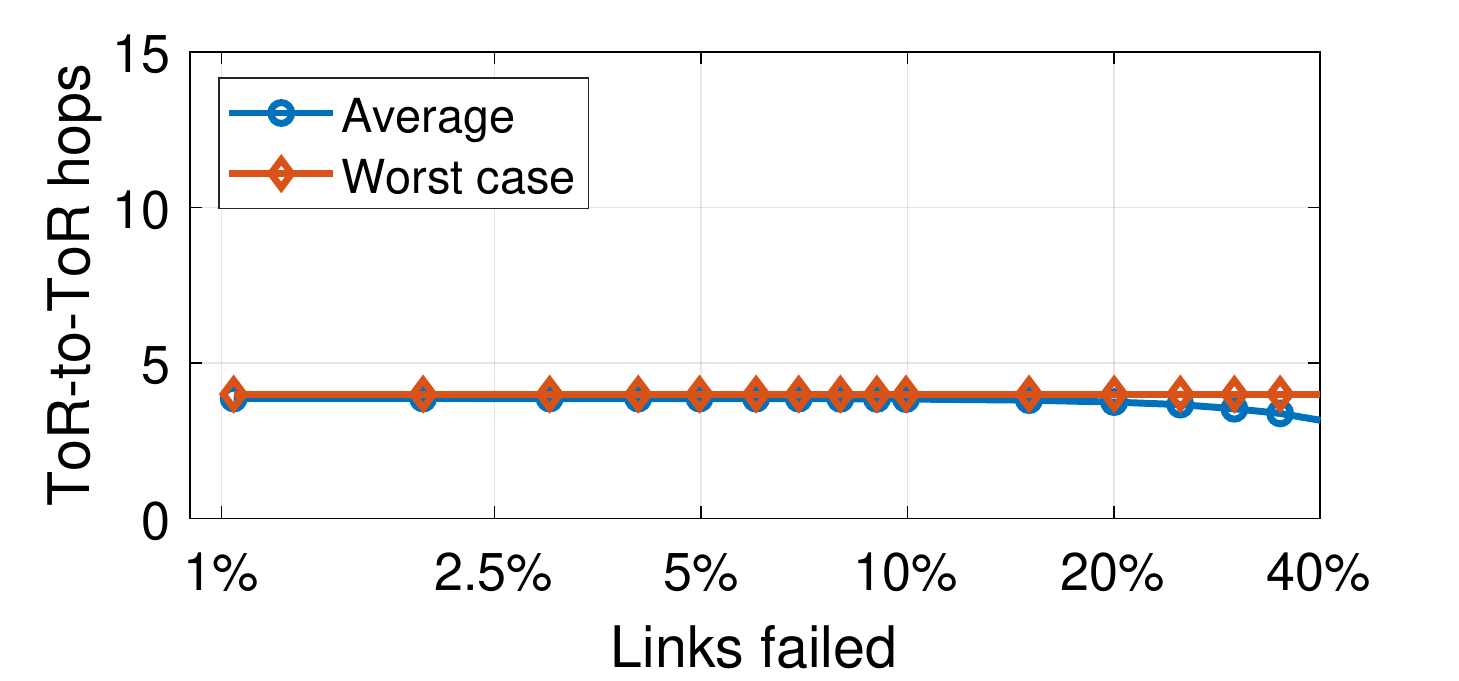} \\
\includegraphics[width=.33\textwidth]{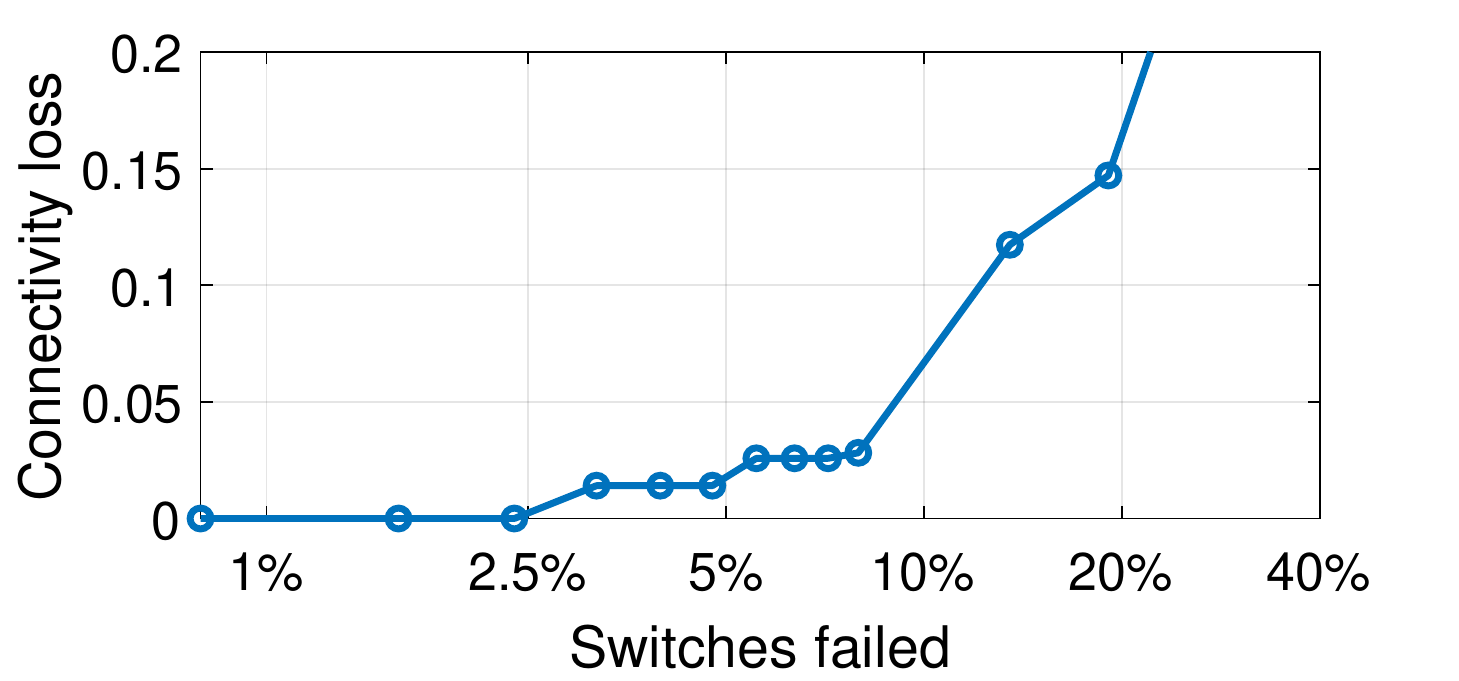}
\includegraphics[width=.33\textwidth]{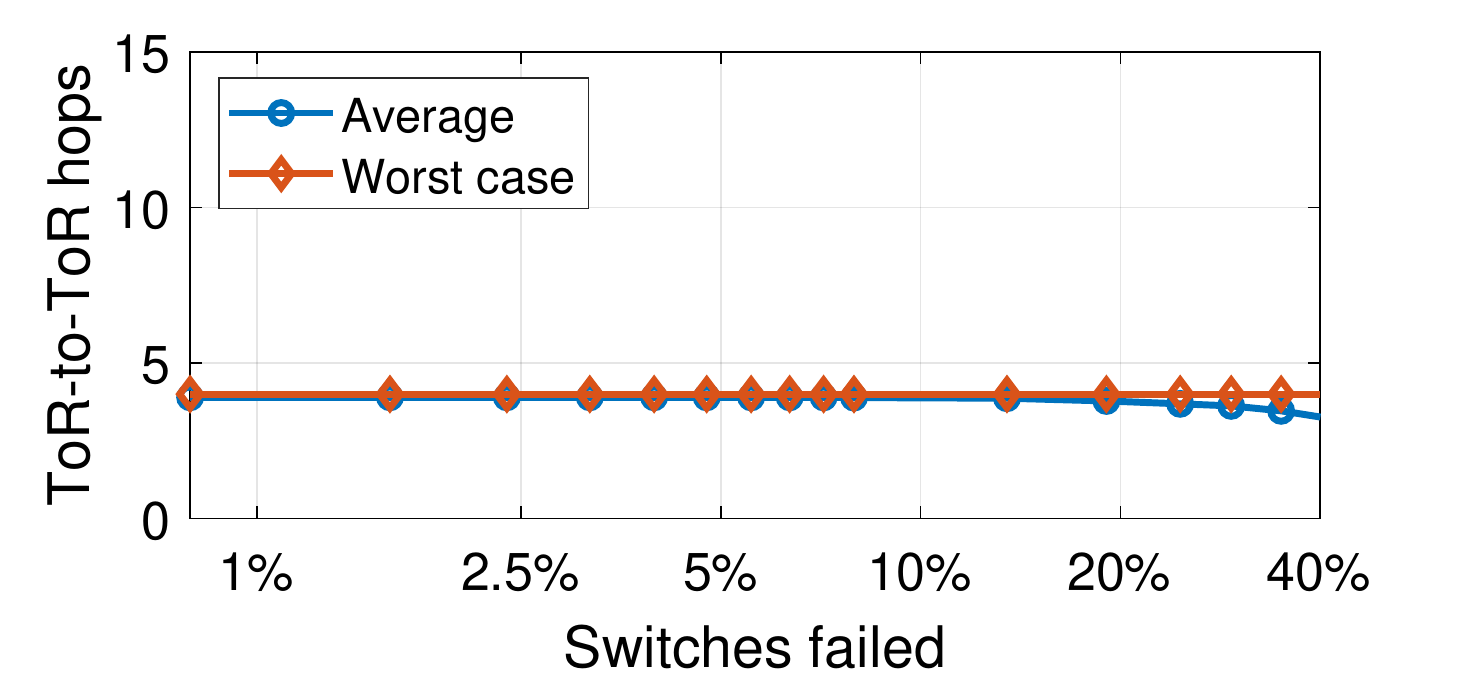}
\caption{\label{fig:ft3to1} Connectivity loss and impact on path lengths in the
3:1 folded Clos for link failures (top two) and ToR failures (bottom two).}
\end{figure*}

\begin{figure*}
\centering
\includegraphics[width=.33\textwidth]{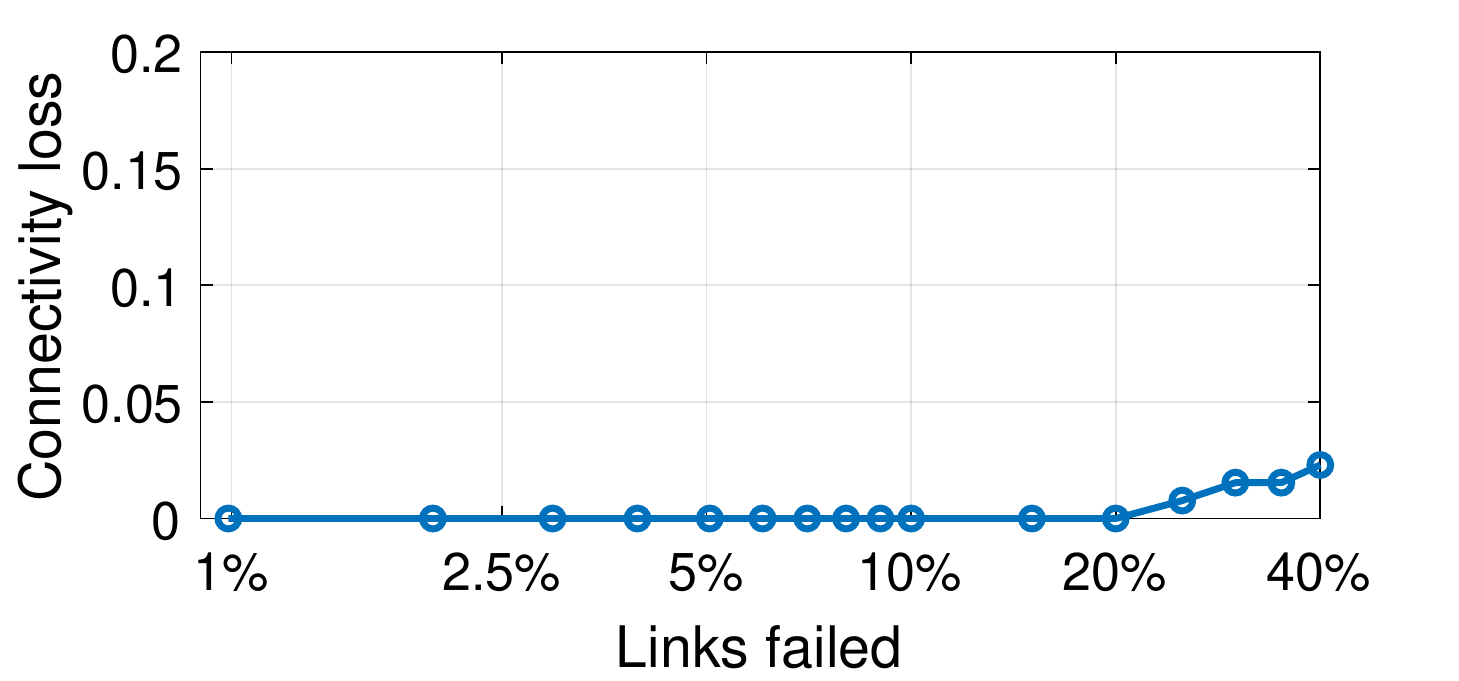}
\includegraphics[width=.33\textwidth]{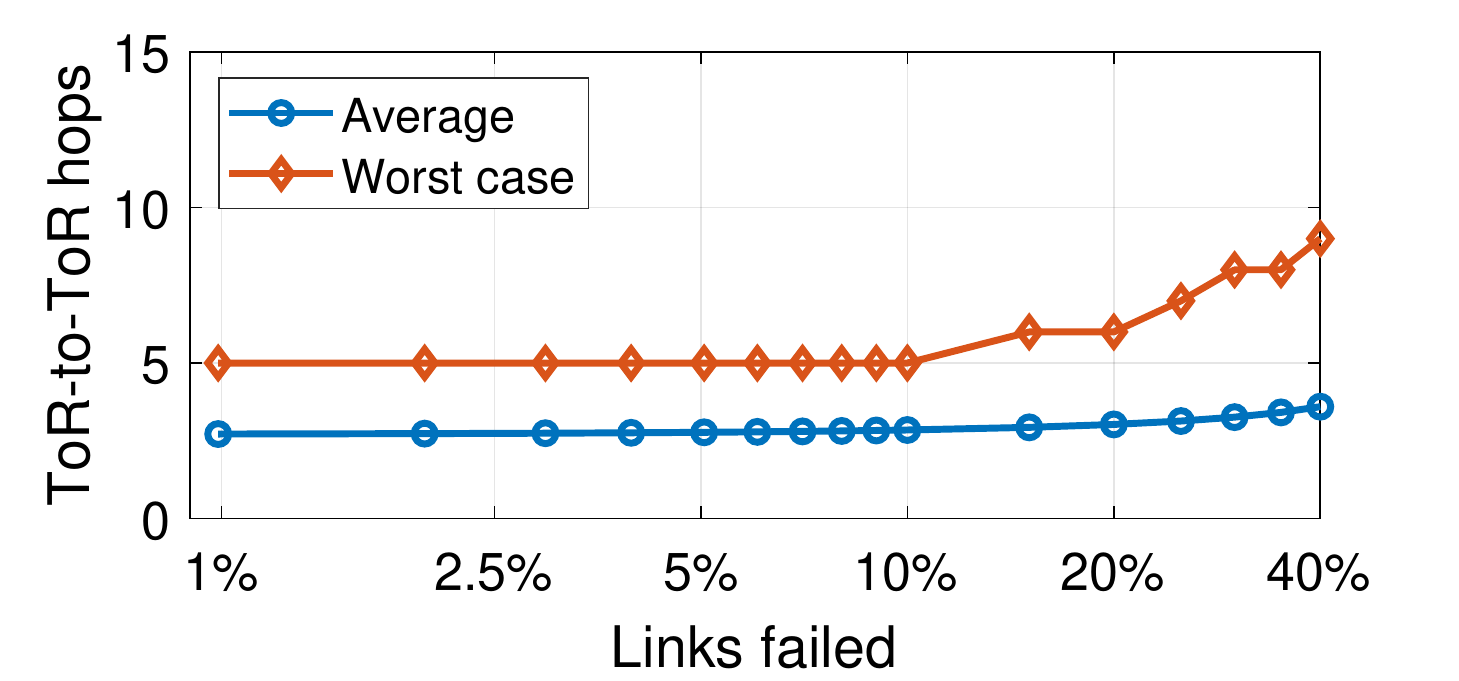} \\
\includegraphics[width=.33\textwidth]{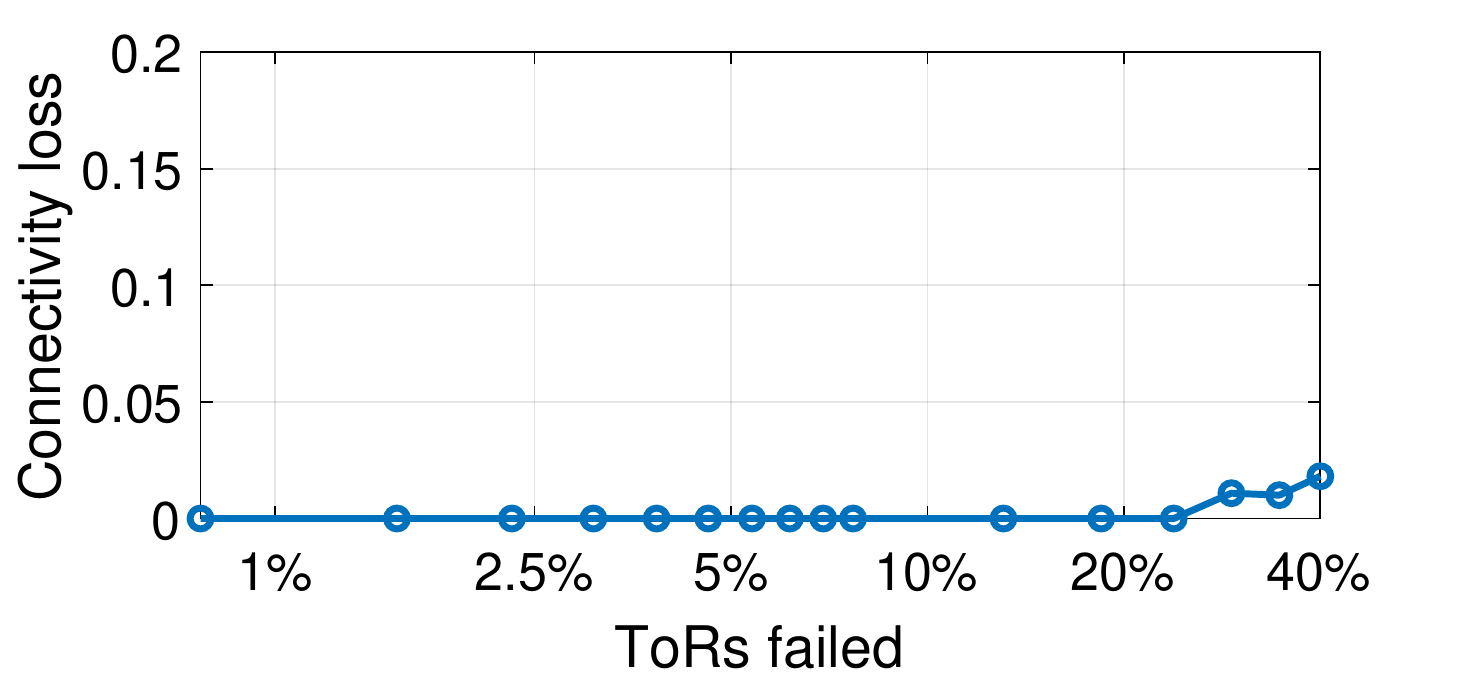}
\includegraphics[width=.33\textwidth]{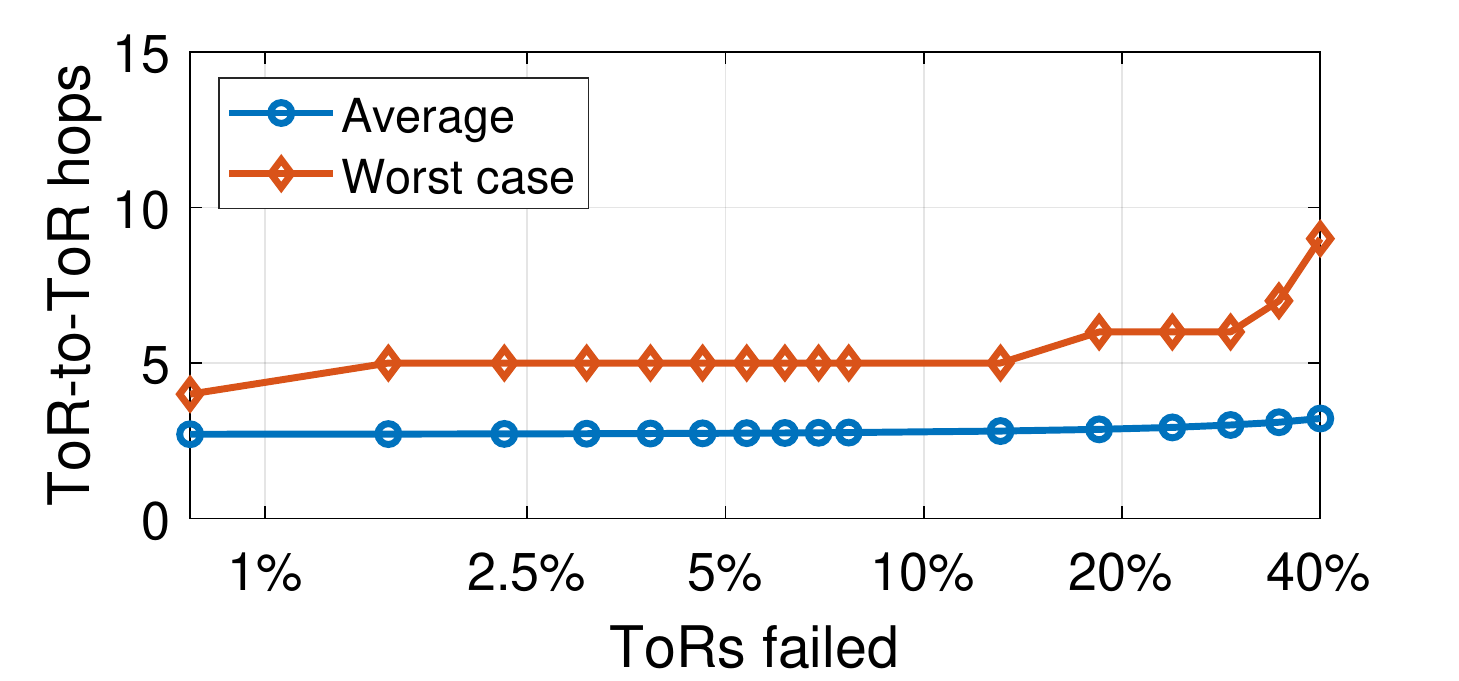}
\caption{\label{fig:ftu7} Connectivity loss and impact on path lengths in the
$u=7$ expander for link failures (top two) and ToR failures (bottom two).}
\end{figure*}

Opera recomputes paths to route around failed links, ToRs, and circuit
switches, and in general these paths will be longer than those under zero
failures.  Figure~\ref{fig:stretch} shows the correlation between the degree of
each type of failure and the average and maximum path length (taken across all
topology slices).

For reference, we also analyzed the fault tolerance properties of the 3:1
folded Clos and $u=7$ expander discussed in the paper. Figure~\ref{fig:ft3to1}
shows the results for the 3:1 Clos and Figure~\ref{fig:ftu7} shows results for
the $u=7$ expander.  We note that Opera has better fault tolerance properties
than the 3:1 folded Clos, but the $u=7$ expander is better yet. This is not
surprising considering the $u=7$ expander has significantly more links and
switches, as well as higher fanout at each ToR.

\end{document}